\documentclass[numberedappendix]{emulateapj}

\shorttitle{Global GMC Evolution}
\shortauthors{Krumholz, Matzner, \& McKee}

\newcommand{\calt}{\mathcal{T}}
\newcommand{\calm}{\mathcal{M}}
\newcommand{\calw}{\mathcal{W}}

\newcommand{\calb}{\mathcal{B}}
\newcommand{\calg}{\mathcal{G}}
\newcommand{\call}{\mathcal{L}}
\newcommand{\cale}{\mathcal{E}}

\newcommand{\drho}{\dot{\rho}}

\newcommand{\cc}{c_{\rm cl}}
\newcommand{\Rc}{R_{\rm cl}}
\newcommand{\dRc}{\dot{R}_{\rm cl}}
\newcommand{\ddRc}{\ddot{R}_{\rm cl}}
\newcommand{\Mc}{M_{\rm cl}}
\newcommand{\dMc}{\dot{M}_{\rm cl}}
\newcommand{\ddMc}{\ddot{M}_{\rm cl}}
\newcommand{\Vc}{V_{\rm cl}}
\newcommand{\Ic}{I_{\rm cl}}
\newcommand{\scl}{\sigma_{\rm cl}}
\newcommand{\gc}{\gamma_{\rm cl}}
\newcommand{\dscl}{\dot{\sigma}_{\rm cl}}

\newcommand{\vkick}{{\mathbf{v}_{\rm ej}'}}
\newcommand{\Sp}{\mathbf{S}_p}
\newcommand{\krho}{k_{\rho}}

\newcommand{\Pa}{P_{\rm amb}}

\newcommand{\Sv}{S_{\rm vir}}
\newcommand{\Vv}{V_{\rm vir}}
\newcommand{\Rv}{R_{\rm vir}}

\newcommand{\sclzero}{\sigma_{\rm cl-0}}

\newcommand{\tcrzero}{t_{\rm cr-0}}
\newcommand{\Rczero}{R_{\rm cl-0}}
\newcommand{\Mczero}{M_{\rm cl-0}}
\newcommand{\etap}{\eta_P}
\newcommand{\etag}{\eta_G}
\newcommand{\etar}{\eta_R}
\newcommand{\azero}{\alpha_{\rm vir-0}}

\newcommand{\etav}{\eta_{\rm v}}
\newcommand{\phiin}{\phi_{\rm in}}
\newcommand{\avir}{\alpha_{\rm vir}}

\newcommand{\nstar}{\mathcal{N}_{*}}

\newcommand{\cs}{c_{\rm s}}
\newcommand{\ci}{a_{\rm I}}
\newcommand{\mphi}{M_{\Phi}}
\newcommand{\etab}{\eta_{\rm B}}

\newcommand{\etae}{\eta_{\rm E}}
\newcommand{\sfrff}{\mbox{SFR}_{\rm ff}}
\newcommand{\tff}{t_{\rm ff}}
\newcommand{\psh}{p_{\rm sh}}
\newcommand{\rsh}{r_{\rm sh}}

\newcommand{\drsh}{\dot r_{\rm sh}}
\newcommand{\cii}{c_{\rm II}}
\newcommand{\PII}{P_{\rm II}}

\newcommand{\msun}{M_{\odot}}

\newcommand{\ltsim}{\protect\raisebox{-0.5ex}{$\:\stackrel{\textstyle <}
        {\sim}\:$}}
\newcommand{\gtsim}{\protect\raisebox{-0.5ex}{$\:\stackrel{\textstyle >}
        {\sim}\:$}}

\slugcomment{Accepted to the Astrophysical Journal, August 10, 2006}

\begin{document}

\title{The Global Evolution of Giant Molecular Clouds\\I:
Model Formulation and Quasi-Equilibrium Behavior}

\author{Mark R. Krumholz\footnote{Hubble Fellow}}
\affil{Department of Astrophysical Sciences, Princeton University,
Princeton, NJ 08544-1001}
\email{krumholz@astro.princeton.edu}

\author{Christopher D. Matzner}
\affil{Department of Astronomy, University of Toronto, Toronto, ON M5S
3H8}
\email{matzner@cita.utoronto.ca}

\author{Christopher F. McKee}\affil{Departments of Physics and
Astronomy, University of California, Berkeley, Berkeley, CA 94720-7304}
\email{cmckee@astron.berkeley.edu}

\begin{abstract}
We present semi-analytic dynamical models for giant molecular clouds
evolving under the influence of HII regions launched by newborn star
clusters. In contrast to previous work, we neither assume
that clouds are in virial or energetic equilibrium, nor do we ignore
the effects of star formation feedback. The clouds, which we treat as
spherical, can expand and contract homologously. Photoionization
drives mass ejection; the recoil 
of cloud material both stirs turbulent motions and leads to an
effective confining pressure. The balance between these effects and
the decay of turbulent motions through isothermal shocks determines
clouds' dynamical and energetic evolution. We find that for realistic
values of the rates of turbulent dissipation, photoevaporation, and
energy injection by HII regions, the massive clouds where most
molecular gas in the Galaxy resides live for a few crossing times, in good
agreement with recent observational estimates that large clouds in
local group galaxies survive roughly $20-30$ Myr. During this time clouds
remain close to equilibrium, with virial parameters of $1-3$ and
column densities near $10^{22}$ H atoms cm$^{-2}$, also in agreement
with observed cloud properties. Over their lives
they convert $5-10\%$ of their mass into stars, after which point most
clouds are destroyed when a large HII region unbinds them. In
contrast, small clouds like those found in the solar neighborhood only
survive $\sim 1$ crossing time before being destroyed.
\end{abstract}


\section{Introduction}

Giant molecular clouds (GMCs) are the primary reservoirs of molecular
gas within spiral galaxies.  Star formation is tightly correlated
with the molecular column density within spiral galaxies \citep{wong02},
and is therefore controlled by the formation and evolution of these
giant clouds. In this and a subsequent paper we develop the theory of
GMC dynamics and present semi-analytical models for GMC evolution.  We
rely on simplifying assumptions about the structure of the cloud and the
properties of the surrounding interstellar medium in order to focus on
clouds' global energy budget and dynamical state.

Observations show that stars form much more slowly than the free-fall
rate in GMCs \citep{zuckerman74,rownd99,wong02,gao04a,wu05}, and 
any successful GMC model must
explain why this should be. It is now widely held that collapse is
inhibited primarily by intensely supersonic motions
\citep{vazquezsemadeni03,maclow04}, rather than magnetic fields
\citep{mouschovias76,shu87,mckee89}. (Star formation is strongly
suppressed in low-extinction regions of molecular clouds, beyond what one would expect if the star formation rate were simply following the column density to some power, suggesting
that magnetic fields may play a secondary role --
\citealt{onishi98,onishi99,onishi02,johnstone04} -- though see
\citealt{hatchell05} who find that low column densities reduce but do not completely prevent star formation.) Simulations and analytic
theory indicate that the observed level of turbulence in GMCs is
sufficient to produce the observed rate of star formation
\citep{krumholz05c}.

However, undriven
supersonic turbulence decays via radiation from isothermal shocks
with an $e$-folding time of roughly one cloud crossing time
\citep{maclow98, stone98, maclow99, padoan99}, so undriven turbulence
alone is not sufficient to prevent global collapse. Instead, GMCs must
either be destroyed before their turbulent motions decay, or the
turbulence must be continually driven. The mode of destruction is
intimately related to the clouds' dynamical state.  Unless a cloud is
destroyed all at once, any internal agent of destruction is also an
internal source for turbulent energy -- and one strong enough to
balance turbulent decay \citep{matzner02}.  Destruction from within
therefore favors models that achieve energetic and dynamical equilibrium
\citep[e.g.][]{mckee89}, if only briefly.  The alternative -- that
clouds are disrupted entirely by outside agents
\citep[e.g.][]{bonnell06b} -- requires most of the cloud
mass to remain gravitationally unbound, as bound regions rapidly
collapse to higher density and become impervious to external
forces. This is very difficult to reconcile with observational
estimates of GMCs lifetimes and ratios of kinetic to
potential energy. We critique the hypothesis of unbound clouds in
greater detail in \S~\ref{discussion}. 

In the following sections, we investigate the properties of molecular
clouds both stirred and destroyed by HII regions within the cloud
volume. The models we present here are improved in several ways relative
to prior work:
\begin{itemize}
\item[-] Rather than enforcing strict mechanical or energetic
  equilibrium, we solve for the time evolution of a cloud's radius and
  turbulent velocity dispersion according to the time-dependent virial
  and energy equations. In an early discussion of this problem, \citet{mckee89}
  allowed for time dependence in the energy equation, but assumed virial equilibrium. \citet{matzner99a} and \citet{matzner99b} followed the evolution of GMCs using time dependent virial and energy equations, but neglected several terms in these equations that we model here. While
our approach is still not a full numerical solution of the equations of
gravity, radiation, and magnetohydrodynamics, this approach enables us to study
cloud dynamics without ignoring the effects of feedback on GMC
  evolution, as most numerical simulations to date have done
  \citep[e.g.][]{clark05}.
\item[-] We account self-consistently for the recoil of cloud matter from
  the sites of mass ejection.  In addition to driving turbulence,
  inward recoil confines the cloud.  Recoil confinement is equivalent to
  an additional (and variable) external pressure, which becomes
  dynamically significant when cloud destruction is
  rapid. Corresponding terms appear in the virial 
  (\S~\ref{equationofmotion}, Appendix \ref{virialderivation}) and energy
  (\S \ref{energyeqn}, Appendix \ref{energyderivation}) equations. 
\item[-] Our model for HII regions (\S \ref{hiiregions}) accounts for the
  scale-dependence of density and velocity structures within GMCs.
  Although this does not fully account for the three-dimensional
  structure of the turbulent cloud medium, it is a significant
  improvement over the uniform cloud model employed by
  \cite{whitworth79}, \cite{williams97}, and \cite{matzner02}.  
\item[-] We apply the \cite{krumholz05c} prescription for the star
  formation rate within turbulent clouds.  This formula accurately
  predicts the star formation rate on a variety of scales, from
  starburst galaxies to the dense precursors of individual star
  clusters \citep{tan06a, krumholz06c}.  We use
  it to govern the birth rate of ionizing associations 
  (\S~\ref{starformation}).
\item[-] Our dynamical simulations (\S~\ref{method}) track the formation
  and evolution of many individual HII regions.  This approach
  accounts for the finite lifetime of ionizing stars and the time
  delay associated with the deceleration of shells driven by HII
  regions, neither of which is negligible compared to GMC dynamical
  times. 
\end{itemize}

We analyze the results of our models in \S~\ref{results},
discuss the implications of our findings in
\S~\ref{discussion}, and summarize our conclusions in
\S~\ref{conclusions}. In a future work (Matzner, Krumholz, \& McKee,
2006, in preparation, hereafter Paper II) we apply this model to the
problem of GMC formation and evolution in the galactic environment,
including the effects of spiral density waves.

We do make several approximations in our work
(\S~\ref{limitations}). We assume that the clouds are spherical and
that they expand or contract homologously. We assume that the clouds
are sufficiently massive that the energy injection is dominated by HII
regions, not protostellar outflows, which based on the models of
\citet{matzner02} should be true for clouds of mass $\sim 10^5$
$\msun$ or more. Such clouds contain most of the molecular mass in the
Galaxy. We neglect the
possibility that the column density of the cloud must exceed a
threshold in order for stars to form \citep[e.g][]{mckee89}. We
neglect energy  injection by HII regions after they reach pressure
equilibrium with the surrounding medium. Finally, we neglect possible
time dependent effects due to the ambient medium of the GMC: no mass
can be added to the cloud, and the ambient pressure remains constant.
Despite these approximations, the models we present in this Paper illustrate the
degree to which GMC properties can be understood in terms of internal
dynamics.

Obviously, real GMCs are not homologously expanding or contracting spheres with smooth density distributions, so the approximations we make to render the problem analytically tractable are quite limiting. For this reason one might be tempted give up on analytic treatment altogether and simply attempt to solve the problem numerically. Unfortunately, full numerical simulation of a giant molecular cloud, including the formation of multiple star clusters and the effects of their feedback, is not feasible with current codes and computers. Instead, one is forced to make numerous approximations regardless of whether one takes a numerical or analytic approach. Many numerical simulations of GMC evolution simply ignore feedback altogether \citep[e.g.][]{clark05}, include it only from a single source and/or focus on size scales much smaller than an entire GMC \citep[e.g.][]{dale05,li06b}, or focus on the galactic scale and lack the resolution to say anything about individual GMCs \citep[e.g.][]{slyz05, tasker06}. For this reason, an analytic approach that allows us to include feedback provides a valuable complement to numerical results, and points out areas for future simulation in which new effects might be discovered by a more thorough treatment of the physics.

\section{Evolution Equations}
\label{basiceqn}

We are interested in the global evolution and energy balance in GMCs,
so we construct a simple model in which we neglect details of cloud
structure. We consider a cloud with density profile $\rho = \rho_e
(r/\Rc)^{-\krho}$, where $\Rc$ is the radius of the cloud edge, and
$\rho_e$ is the edge density. As we discuss below, we take $\krho=1$
as typical of GMCs. We approximate that the cloud evolves
homologously, so that $\krho$ is constant. However, the cloud can
expand or contract and can lose mass (via evaporation of gas by HII
regions), so $\rho_e$ and $\Rc$ both vary in time. The cloud is
embedded in an ambient medium of negligible density and constant
pressure $\Pa$. We model evaporation of gas from the cloud as a
wind, into which cloud material is injected at a rate $\drho$ (which
we take to be negative). Gas that is injected into the wind travels
radially outward with ``kick'' velocity $\vkick$ relative to the
radial velocity of the cloud at that radius. Homology requires that
the mass loss rate follow the existing density profile, so
\begin{equation}
\drho = \frac{\dMc}{\Mc} \rho,
\end{equation}
where $\Mc$ is the total mass of the cloud. We assume that the wind is
low density and escapes from the vicinity of the cloud quickly, so
that we can neglect its gravitational interaction with the cloud. As
we discuss in more detail in \S~\ref{hiiregions}, this is a reasonable
model for mass evaporating from an HII region.

We neglect the possibility that turbulent motions within the cloud are likely to lead to a significant loss of mass. This seems justified, since in a GMC with a virial ratio of unity, roughly the value for observed GMCs, the 3D turbulent velocity dispersion is smaller than the escape speed from the GMC surface by a factor of $\sim 2$. Since the distribution of velocities in a supersonically turbulent medium cuts off exponentially above the turbulent velocity dispersion \citep{krumholz06a}, there is a negligible amount of mass moving rapidly enough to escape. We also neglect the possibility that a GMC might gain mass during its evolution, due to continuing infall. This assertion is more problematic, and we discuss it in more detail in \S~\ref{limitations}.

Within the limitations of these assumptions, we derive the
Eulerian Virial Theorem (EVT) and equation of energy conservation in
Appendices \ref{virialderivation} and \ref{energyderivation}. We then
use these to construct evolution equations for the the cloud.

\subsection{Equation of Motion}
\label{equationofmotion}

We derive the equation of motion from the EVT for an evaporating
homologously-moving cloud,
\begin{eqnarray}
\frac12 \ddot{I}_{\rm cl} & = & 
2(\calt-\calt_0) + 
\calb+\calw-\frac12
\frac{d}{dt}\int_{\Sv} (\rho\mathbf{v} r^2)\cdot d\mathbf{S} 
\nonumber\\ 
& & {} +
2 \ci \dMc \Rc \dRc + \frac12 \ci \ddMc \Rc^2
\nonumber\\
& & {} +
 \frac{3-\krho}{4-\krho} \dMc \Rc v_{\rm ej}'.
\label{EVTtext}
\end{eqnarray}
The proof of this theorem is in Appendix \ref{virialderivation}.
In this equation $I_{\rm cl}$ is the cloud moment of inertia, $\calt$ is
the total kinetic and thermal energy, $\calt_0$ is the energy
associated with the confining external pressure, $\calb$ and $\calw$
are the magnetic and gravitational potential energies, and the surface
integral represents the rate of change of the flux of inertia across
the surface $\Sv$ that bounds the volume to which we apply the virial
theorem. These are all terms that appear in the EVT for a
cloud without a wind, and their formal definitions are given in the
Appendix. The three additional terms represent the second derivative
of cloud inertia caused by mass loss through the wind (the first two
extra terms) and the rate at which recoil from the process of
launching the wind injects momentum into the cloud (the final
additional term).

We now evaluate each of these terms in the context of our model.
The moment of inertia is
$I_{\rm cl}=\ci \Mc \Rc^2$, where $\ci\equiv (3-\krho)/(5-\krho)$, 
so its second derivative is
\begin{eqnarray}
\frac{1}{2}\ddot{I}_{\rm cl} & = &
\ci \Mc \dRc^2 + \ci \Mc \Rc \ddRc
\nonumber\\
& & \: {} +
 2 \ci \dMc \Rc \dRc + \frac{1}{2}
\ci \ddMc \Rc^2.
\end{eqnarray}
Next consider the kinetic term $\calt$, which we evaluate by
decomposing the velocity into large-scale homologous and
fluctuating turbulent components (equation \ref{vdecomp}). This gives
\begin{equation}
\calt =\frac{3}{2} \Mc \cc^2+\frac{1}{2} \ci \Mc
\dRc^2 + \calt_{\rm turb} + 2\pi\Pa (\Rv^3-\Rc^3),
\end{equation}
where $\cc$ is the sound speed in the cloud (assumed constant),
$\calt_{\rm turb}$ is the term for turbulent motions, and the last
term comes from the constant ambient pressure $\Pa$ outside the
cloud. We return to $\calt_{\rm turb}$ below. Note that our assumption
of homologous motion
implicitly neglects the possibility of significant rotational
motions. Observed GMCs have negligible kinetic energies in overall
rotation compared to turbulent motions or gravitational potential
energy.

We use the same strategy for the gravitational and magnetic terms as
for the kinetic term, dividing them into steady and fluctuating parts.
The non-turbulent gravitational part is
\begin{equation}
\calw_{\rm non-turb} = -\frac{3}{5} a \frac{G \Mc^2}{\Rc},
\end{equation}
where 
\begin{equation}
a = \frac{15-5 \krho}{15-6 \krho}.
\end{equation}
In principle we should include a component of potential energy due to
stars, but all observed molecular clouds have gas
masses that greatly exceed their stellar masses. For this reason, we
may neglect the stellar mass.
For the non-turbulent magnetic component, we follow \citet{mckee93}. Let
$\Phi$ be the total magnetic flux threading the cloud, so that the
mean field within the cloud is $\overline{B}=\Phi/(\pi
\Rc^2)$. From the the form of the magnetic energy term (equation
\ref{calbdef}), it is clear that the non-turbulent component of this
term scales as $\calb_{\rm non-turb}
\propto \overline{B}^2 \Rc^3$. We therefore define the constant $b$
such that
\begin{equation}
\calb_{\rm non-turb} = \frac{b}{3} \overline{B}^2 \Rc^3 =
\frac{b}{3 \pi^2} \left(\frac{\Phi^2}{\Rc}\right).
\end{equation}
The exact value of $b$ depends on the topology of the magnetic field
and on the background field $B_0$, but it is generally of order
unity, with $b=0.3$ as a typical value \citep{mckee93}. We now define
the magnetic critical mass $\mphi$ by 
\begin{equation}
\mphi^2 = \left(\frac{5b}{9\pi^2 a}\right) \frac{\Phi^2}{G},
\end{equation}
so that we have
\begin{equation}
\calb_{\rm non-turb} = \frac{3}{5} a \frac{G \mphi^2}{\Rc}.
\end{equation}
We define the magnetic support parameter by
\begin{equation}
\etab = \frac{\mphi}{\Mc}.
\end{equation}
With this definition, we can combine the non-turbulent magnetic and
gravitational terms to find
\begin{equation}
\calw_{\rm non-turb}+\calb_{\rm non-turb} =
-\frac{3}{5} a (1-\etab^2) \frac{G \Mc^2}{\Rc}.
\end{equation}

Now consider the turbulent components. First, we can neglect the
turbulent component of the gravitational term because since most
sub-regions of a molecular cloud are not self-gravitating
\citep{krumholz05c}, and therefore have negligible potential energy in
comparison to their kinetic or magnetic energies. We can therefore set
$\calw=\calw_{\rm non-turb}$. For the magnetic and
kinetic turbulent components, \citet{mckee92} argue for equipartition
of kinetic and magnetic energy. \citet{stone98} find in simulations of
low plasma-$\beta$ turbulence that magnetic energy is slightly
sub-equipartition, $\calb_{\rm turb} \approx 0.6\, \calt_{\rm turb}$,
which \citet{mckee03} argue can be understood as the kinetic and
magnetic energies reaching equipartition for motions transverse to the
field, but not for motions along the field. We adopt the ratio of
magnetic to kinetic energy found by \citet{stone98}, so the combined
turbulent kinetic and magnetic energies in the virial theorem are
\begin{equation}
2\calt_{\rm turb} + \calb_{\rm turb} \approx
2.6\, \calt_{\rm turb} = 3.9\, \Mc \scl^2,
\end{equation}
where 
$\scl=\left\langle (v_z-v_{{\rm cl},z})^2\right\rangle^{1/2}_{\rho}$
is the 
one-dimensional mass-weighted turbulent velocity dispersion of the
gas in the cloud.

Finally, we come to the surface terms, $\calt_0$ and $(d/dt)\int_{\Sv}
(\rho\mathbf{v}r^2)\cdot d\mathbf{S}$. We can make the latter term
zero by choosing our virial surface to be well outside the cloud so
that the density of cloud material on the surface is negligible. However,
the pressure outside the cloud $\Pa$ is non-zero, so
\begin{equation}
\calt_0 = 2\pi\Pa\Rv^3.
\end{equation}

Substituting into the EVT (\ref{EVTtext}), we arrive at an equation of
motion for the cloud:
\begin{eqnarray}
\ci \ddRc & = & 
3.9\frac{\scl^2}{\Rc} + 3\frac{\cc^2}{\Rc}
- \frac{3}{5} (1-\etab^2) a \frac{G \Mc}{\Rc^2} 
\nonumber \\
& & \: {} - 
4\pi \Pa
\frac{\Rc^2}{\Mc}
+ \left(\frac{3-\krho}{4-\krho}\right) \frac{\dMc}{\Mc} v'_{\rm ej}.
\label{virialdim}
\end{eqnarray}
This equation is intuitively easy to understand. The left-hand side
represents the acceleration of the cloud edge, which is equated with
the force per unit mass due to internal turbulence and pressure (the
first two terms), gravity and magnetic fields (the third term),
external pressure (the fourth term), and recoil from the evaporating
gas (the final term).

For convenience we wish to non-dimensionalize this equation. Let
$\Mczero$, $\Rczero$, and $\sclzero$ be the initial mass, radius, and
velocity dispersion of the cloud. We define the dimensionless variables
$M=\Mc/\Mczero$, $R=\Rc/\Rczero$, $\sigma=\scl/\sclzero$, and
$\tau=t/\tcrzero$, where $\tcrzero=\Rczero/\sclzero$ is the
crossing time of the initial cloud. In these variables
(\ref{virialdim}) becomes
\begin{equation}
\label{eqofmotionfinal}
R'' =
\frac{3.9\, \sigma^2 + 3 \calm_0^{-2}}{\ci R}
-\etag \frac{M}{R^2} - \etap \frac{R^2}{M} + \etar \frac{M'}{M} 
\end{equation}
where the primes indicate differentiation with respect to $\tau$,
\begin{eqnarray}
\label{etageqn}
\etag & \equiv &
\frac{3 a (1 - \etab^2)}{\ci \azero}, \\
\label{etapdefn}
\etap & \equiv &
\frac{4 \pi \Rczero^3 \Pa}{\ci \Mczero \sclzero^2}, \\
\label{etardefn}
\etar & \equiv &
\left(\frac{3-\krho}{4-\krho}\right)
\frac{v'_{\rm ej}}{\ci \sclzero}
\end{eqnarray}
and we have defined the initial Mach number
\begin{equation}
\calm_0 \equiv \frac{\sclzero}{\cc}
\end{equation}
and non-thermal virial parameter \citep{bertoldi92}
\begin{equation}
\azero \equiv \frac{5\sclzero^2 \Rczero}{G \Mczero}.
\end{equation}
If the cloud's initial state is in equilibrium and it is not losing mass, 
so $R'(0)=M'(0)=0$,
then the ambient pressure $\Pa$ must be such that
\begin{equation}
\label{etapeqn}
\etap = \frac{3.9+3\calm_0^{-2}}{\ci}-\etag.
\end{equation}
Note that this will generally 
implies an ambient pressure higher than the mean in the ISM.  We
consider it realistic, however, because GMCs form in relatively
overpressured regions, and because we have not included the weight of
an atomic layer overlying the cloud.  Once mass loss commences, there
will be an additional recoil pressure. 

\subsection{Equation of Energy Evolution}
\label{energyeqn}

To derive the evolution equation for the cloud energy, we begin with
the general energy conservation equation for an evaporating homologous
cloud, which we derive in Appendix \ref{energyderivation} and for
convenience repeat here:
\begin{eqnarray}
\frac{d\cale}{dt} & = & \frac{\dMc}{\Mc} \left[\cale +
(1-\etab^2) \calw\right] - 4\pi\Pa\Rc^2\dRc
\nonumber \\
& & \: {} +
\left(\frac{3-\krho}{4-\krho}\right) \dMc \dRc v'_{\rm ej} +
\calg_{\rm cl} - \call_{\rm cl}.
\label{energyeqn1text}
\end{eqnarray}
Here $\cale$ is the total cloud energy, and $\calg_{\rm cl}$ and
$\call_{\rm cl}$ are the rates of radiative energy gain and loss
integrated over the entire cloud. This equation is easy to understand
intuitively. The term $(\dMc/\Mc) \cale$ is  simply the mass loss rate
times the energy per unit mass in the  cloud. The term
$(\dMc/\Mc) (1-\etab^2)\calw$ is the rate at which mass loss reduces
the cloud energy via reduction of the gravitational and magnetic
fields, both of which are proportional to the mass. The next two terms
represent the rate at which the external pressure and the recoil force
from launching the wind do work on the cloud. Finally,  the last two
terms are simply the rate of radiative gains and losses. 

Using the same arguments as in \S~\ref{equationofmotion}, we may write
the total energy in the cloud as
\begin{equation}
\label{evircl}
\cale = \frac{1}{2} \ci \Mc \dRc^2 + 
2.4\, \Mc \scl^2
+\frac{3}{2} \Mc \cc^2 - \frac{3}{5} a (1-\etab^2) \frac{G
\Mc^2}{\Rc}.
\end{equation}
Note that the factor of $2.4$ in the $\Mc \scl^2$ term comes from
taking $\calt_{\rm turb}+\calb_{\rm turb} \approx 1.6\, \calt_{\rm
turb}$, and the $3/2$ in front of the $\Mc \cc^2$ term comes from the
assumption that the ratio of specific heats for the cloud is
$\gc=5/3$. One might expect $7/5$ instead, since the cloud is
molecular and therefore diatomic. However, the lowest rotational or
vibrational excitations of H$_2$ have excitation temperatures of
several hundred K. Since the cloud is far colder than this, molecules
never have enough energy to access their rotational and vibrational
degrees of freedom, and the gas acts as if it were monatomic. The time 
derivative of this is
\begin{eqnarray}
\frac{d\cale}{dt} & = &
\frac{1}{2}\ci \dMc \dRc^2 + \ci \Mc \dRc \ddRc +
2.4\, \dMc \scl^2
\nonumber \\
& &
{} + 4.8\, \Mc \scl \dscl
\frac{3}{2} \dMc \cc^2
- \frac{6}{5} a (1-\etab^2) \frac{G \Mc
\dMc}{\Rc} 
\nonumber \\
& &
{} + \frac{3}{5} a (1-\etab^2) \frac{G \Mc^2 \dRc}{\Rc^2}.
\label{dedtvircl}
\end{eqnarray}
Substituting $\cale$ and $d\cale/dt$ into the energy equation
(\ref{energyeqn1text}), we find
\begin{eqnarray}
\lefteqn{\ci \Mc \dRc \ddRc + 4.8\Mc \scl \dscl}
\qquad
\nonumber \\
\lefteqn{
{} + 
\frac{3}{5} a (1-\etab^2) \frac{G \Mc^2}{\Rc^2} \dRc 
=} \qquad
\nonumber \\
& &
-4 \pi \Pa \Rc^2 \dRc +
\left(\frac{3-\krho}{4-\krho}\right) \dMc \dRc v'_{\rm ej} 
\nonumber \\
& & \; {}
+
\calg_{\rm cl} - \call_{\rm cl}.
\end{eqnarray}
We may regard this as an evolution equation for $\scl$, which makes
intuitive sense: the overall expansion and contraction of the cloud is
dictated by the equation of motion, and the thermal energy per unit
mass is fixed,
so the turbulence acts as the energy resevoir, increasing or
decreasing as the cloud gains or loses energy. Re-arranging to solve
for $\dscl$ and non-dimensionalizing as we have done with the equation
of motion gives
\begin{equation}
\label{energyeqfinal}
\frac{4.8}{\ci} \sigma' = -\frac{R' R''}{\sigma}
- \etag \frac{M R'}{R^2 \sigma}
- \etap \frac{R^2 R'}{M\sigma}
+ \etar \frac{M' R'}{M \sigma}
+ \frac{\calg - \call}{\ci M \sigma},
\end{equation}
where $\calg$ and $\call$ are the dimensionless rates of radiative
energy gain and loss, defined by
\begin{equation}
\calg - \call = \left(\frac{\Rczero}{\Mczero\sclzero^3}\right)
(\calg_{\rm cl} - \call_{\rm cl}).
\end{equation}

\section{Energy Sources and Sinks}
\label{sources}

In this section we evaluate the rates of radiative energy gain $\calg$
and loss $\call$, and the characteristic launch speed of evaporating gas
$v'_{\rm ej}$. Together with the equation of motion
(\ref{eqofmotionfinal}) and the energy equation (\ref{energyeqfinal}),
and the star formation rate (\S~\ref{starformation}),
this will completely specify the evolution of our model clouds.

\subsection{Decay of Turbulence via Isothermal Shocks}
\label{turbdecay}

GMCs are approximately isothermal because their radiative time scales
are much shorter than their mechanical time scales. As a result,
supersonic motions within the cloud generate radiative shocks that
remove energy from the cloud. The
problem of the decay of turbulent motions by supersonic isothermal
shocks in both hydrodynamic and magnetohydrodynamic media has been
studied extensively by numerical simulation \citep[e.g.][]{maclow98,
stone98, maclow99, padoan99}. \citet{stone98} finds that the
dissipation time scale of the turbulent energy is $t_{\rm dis} \equiv
\dot{E}/E = 0.83\, \lambda_{\rm in}/v_{\rm rms} = 0.48 \lambda_{\rm
in}/\sigma$, where $\lambda_{\rm in}$ is the length scale on which the
energy is injected.

In reality, HII
regions coming from associations of various sizes, winds, and
gravitational contraction of the cloud will all contribute to
turbulent motions and inject energy on different length
scales. However, we can estimate an effective length scale through a
combination of observational and theoretical
considerations. Observationally, turbulence in nearby GMCs appears to
be driven on scales comparable to the cloud scale or larger
\citep{basu01, heyer04}, and theoretical estimates of the effective
energy injection scale of HII regions suggest that this is also near
the cloud scale \citep{matzner02}. The longest wavelength mode that
a cloud of radius $\Rc$ can support is $\sim 4 \Rc$, where the factor
of four arises because the largest turbulent mode corresponds to
overall expansion or contraction of the cloud, in which diametrically
opposed points are moving in opposite directions. Motion in
opposite directions corresponds to the points being half a wavelength
apart, giving a total wavelength of twice the cloud diameter
\citep{matzner02}. Thus, we take the effective injection scale to be
$\lambda_{\rm in} = 4\phi_{\rm in} \Rc$, where $\phi_{\rm in}\leq 1$,
and we take $\phi_{\rm in}=1$ as a fiducial value based on
observations and theory. The (dimensionless) rate at which energy is
radiated away due to decaying turbulence is
\begin{equation}\label{Lambda_diss}
\call = \frac{\etav}{\phi_{\rm in}} \frac{M \sigma^3}{R},
\end{equation}
where Stone et al.'s simulations give $\etav=1.2$. Note that, in
deriving this factor, we have used our result that the energy in the
turbulence, including magnetic and kinetic contributions, is $2.4\,
\Mc \scl^2$.

It is also worth noting the possibility that the measured energy loss
rates are too high. \citet{cho03} argue that Alfv\'en waves cascade
and decay anisotropically, and that this anisotropy can reduce the
decay rate. However, Cho \& Lazarian argue that simulations to date
have not captured this effect because they use an isotropic
driving field that is unrealistic. \citet{sugimoto04}
simulate filamentary molecular clouds, and find that Alfv\'en waves of
certain polarizations decay more slowly than simulations have
found. Even if neither of these effects apply, there are differences
in the rate of decay in different simulations depending on how the
turbulence is driven. The simulations of \citet{maclow99} give
slightly lower dissipation rates, probably because in those
simulations the turbulence is forced with a driving field that is
constant in time, while in those of \citet{stone98} the driving field
is determined randomly at each time step. While we feel that the
Stone et al. approach is somewhat more realistic, this is by no
means certain.

\subsection{HII Regions}
\label{hiiregions}

HII regions are the dominant source of energy injection into GMCs
from within \citep{matzner02}.  We consider the effects of the 
HII region from a single association here, and
extend our results to a population of associations to
\S~\ref{starformation}.

\subsubsection{Evolution of Individual HII Regions}
\label{HIIevol}

We first describe the evolution of an HII region embedded in a
molecular cloud, modifying the analysis by \cite{matzner02} by
allowing the mean ambient density and velocity dispersion to vary with
radius $r$ away from the formation site of an association, as
$\rho(r)\propto r^{-\krho}$ and
$\sigma(r)\propto r^{-k_\sigma}$, respectively.  \cite{matzner02}
considered only the homogeneous case $\krho = k_\sigma=0$.  Note that
the local turbulent virial parameter $\alpha(r)\equiv 5
r\sigma^2(r)/[G M(r)]$ scales as $r^{\krho -2k_\sigma -2}$.  Since
$\alpha(r)$ is roughly unity on the scale of the cloud and on that of
the newborn association (else it would not have formed), it 
is reasonable to assume $\krho = 2k_\sigma+2$.  The observed
line width-size relation and density-size relations \citep{solomon87} imply
$k_\sigma=-1/2$ and $\krho=1$, and we adopt these values below. These
parameters correspond to a cloud with a negligible internal pressure
gradient. Note that it has been suggested that the observational result $\krho=1$ (which is equivalent to GMCs having constant column densities within a galaxy) is simply an observational artifact \citep[e.g.][]{vazquezsemadeni97}. However, many of the proposed mechanisms to explain how this artifact could be created do not apply to extragalactic observations, and more recent observations show that GMCs in other galaxies also show constant column densities \citep[and references therein]{blitz06a}. We discuss this point in more detail in \S~\ref{varcolsection} and \S~\ref{gmcevol}, and also refer readers to the discussion of this point in \citet{krumholz05c}.

The density variation affects the expansion phase of the HII region,
and the variation in velocity dispersion affects how HII regions merge
with the background turbulence.  
The evolution
of an HII region in a turbulent GMC is a substantial problem that must
ultimately be solved via simulation. However, to date only preliminary
attempts to solve the problem have been made
\citep[e.g.][]{dale05,mellema05,maclow06,krumholz06e}, so we are forced to rely an analytic
approximations.

First, consider the expansion phase of an HII region.  Assume
that it has expanded well beyond its initial
Str\"omgen radius. The mean ambient density is $\bar{\rho}(r)\propto
r^{-k_\rho}$ within a distance $r$ from the association, the ionized
gas temperature is $T_{\rm II}\simeq 7000$ K, the (constant) ionizing
luminosity is $S=10^{49}S_{49}$ s$^{-1}$, and the recombination
coefficient is $\alpha^{(2)}$ in the on-the-spot approximation. If the
HII region is blister-type, ionized gas will rocket away from the
ionization front at a velocity $v'_{\rm ej} = 2 c_{\rm II}$, where
$c_{\rm II} = 9.74$ km s$^{-1}$ is the sound speed in the ionized
gas. This is the characteristic launch velocity of our cloud's
escaping wind. For the evolution of the HII region,
\citeauthor{matzner02}'s dynamical equation [his eq. (15)] admits the
self-similar solution
\begin{equation}
r_{\rm sh}^2 \bar{\rho}(r_{\rm sh}) = (2,1)\times
\frac{3}{4}\frac{(7-2\krho)^2}{9-2\krho} \rho_{\rm II} c_{\rm II}^2 t^2
\end{equation}
for (blister, embedded) regions, respectively. Here $r_{\rm sh}$ is
the radius of the shell at time $t$, and $\rho_{\rm II}$ is the
(uniform) density inside the HII region, which must vary as $r_{\rm
sh}^{-3/2}$ to ensure that ionizations balance recombinations. This
equation applies when the expansion velocity is much larger than the
turbulent velocity dispersion; we discuss the effects of turbulence
below. This yields 
\begin{equation}\label{HIIselfsim}
r_{\rm sh}^{7/2}\bar{\rho}(r_{\rm sh}) = (2,1)
\times 2.2 \frac{3(7-2\krho)^2}{2(9-2\krho)}k_B T_{\rm II} 
\left(\frac{3S}{4\pi\alpha^{(2)}}\right)^{1/2} t^2 
\end{equation} 
where $k_B$ is Boltzmann's constant.  The leading factor of 2.2
accounts for the particle abundances assuming helium is singly
ionized.  We assume blister-style regions hereafter, because GMCs are
porous and therefore even small HII regions are likely to be able to
punch through to low density channels and vent \citep{matzner02}. For
our fiducial case $\krho=1$, the mean column $\Sigma\equiv
4\bar{\rho}(r) r/3 $ is a constant. Adopting $\alpha^{(2)} =
3.46\times 10^{-13}$ cm$^3$ s$^{-1}$, and adjusting the effective
ionizing luminosity downward by a factor 1.37 to account for dust
absorption \citep{williams97}, we find
\begin{equation} \label{rdrv} 
r_{\rm sh} = 15.4 S_{49}^{1/5} 
\left(\frac{t}{3.8\;\rm Myr}\right)^{4/5}
\left(\frac{0.03\;\rm g\;cm^{-2}}{\Sigma}\right)^{2/5}{~\rm pc}.
\end{equation} 
Here we have normalized $\Sigma$ to a typical value for Milky Way
molecular clouds.  We take the typical ionizing lifetime to be 3.8 Myr
because we adopt the \cite{kroupa01a} initial mass function (IMF) and
the stellar ionizing luminosities and lifetimes tabulated by
\cite{parravano03}.  
However, the lifetime can fluctuate from this for small clusters, so
we define $t_{\rm ms}$ as the main-sequence ionizing lifetime of 
stars in a given cluster.
(With these choices, a young association large enough to fully sample
the IMF emits $3.4\times 10^{46}$ ionizing photons per second per
solar mass, and has one star above $8\msun$, hence one supernova, per
131 solar masses.)  The momentum of the HII region during this driving
phase is
\begin{eqnarray}
p_{\rm sh} & = & 5.1\times 10^{43} S_{49}^{3/5} 
\left(\frac{t}{3.8\;\rm Myr}\right)^{7/5}
\nonumber \\
& & {} \times
\left(\frac{0.03\;\rm g\;cm^{-2}}{\Sigma}\right)^{1/5}{\rm
  g\;cm\;s^{-1}}.
\label{pdrv}
\end{eqnarray}
Assuming the gas can escape freely, the mass evaporation rate from the
cloud is simply 
\begin{equation} \label{Mev}
\dMc = -\dot{p}_{\rm sh}/(2c_{\rm II}).
\end{equation} 
Note that equations (\ref{rdrv}) and (\ref{pdrv}) are quite
insensitive to the actual escape of ionized gas \citep{matzner02}, but
$\dMc$ depends on the existence of an escape route. 

After the driving stars burn out at $t=t_{\rm ms}$
(and if it has not yet merged with the cloud turbulence,
\S~\ref{HII-merging}), the HII region will continue to expand in a
momentum-conserving snowplow, in which $\dot{r}_{\rm sh}\propto
r_{\rm sh}^{-(3-\krho)} \rightarrow r_{\rm sh}^{-2}$ so that
\begin{equation}
r_{\rm sh} = \left[r_{\rm ms}^{3} + 3 r_{\rm ms}^2 \dot{r}_{\rm ms}
(t - t_{\rm ms}) \right]^{1/3},
\end{equation}
where $r_{\rm ms}$ and $\dot{r}_{\rm ms}$ are the radius and velocity
of the expanding shell at the time when the driving stars burn out.
No mass is evaporated in this phase; we ignore the possibility of
dynamical mass ejection by the snowplow unless the snowplow contains
enough kinetic energy to unbind the cloud as a whole.

We now comment on two potential criticisms of our HII
regions.
First, real molecular clouds are very inhomogeneous, being filled with
clumpy and filamentary structure.  To what degree should this affect
the results of this section?  Very little, we argue.
\citet{mckee84} consider HII regions in a medium composed
entirely of dense clumps.  Adopting a clump mass distribution very
similar to that observed within molecular clouds, they find (their
eq.\ 5) that the photoevaporative rocket effect clears all but a
couple clumps from within $\rsh(t)$.  Additional homogenization, not
considered by \citeauthor{mckee84}, should come from the
ram pressure stripping of overdense clumps and filaments as they are
struck by gas already set in motion.  For these reasons we expect the
gross properties of HII regions to be rather insensitive to local
inhomogeneities.  Recent simulations by \citet{mellema05} and
\citet{maclow06} support this expectation, and \citeauthor{mellema05}
find that the radius of an HII region in a turbulent medium as a
function of time is very similar to what one would find in a uniform
medium. (The simulation by \citet{dale05} gives a somewhat different
result, but its different initial conditions and brief duration do not
allow a fair comparison. In particular, \citeauthor{dale05} compare
their HII region expansion rate to the analytic solution for a uniform
medium, despite the fact that they are simulating a strongly centrally
condensed gas cloud.)  For these reasons we
expect our results to be robust against clumping of the gas on scales
smaller than $\rsh(t)$.  Given that our adoption of blister-type flow
reflects the existence of inhomogeneities on scales larger than
$\rsh(t)$, we expect the results in this section to be robust against
cloud inhomogeneities in general.

A second potential criticism concerns our assumption that each HII
region expands as if the gas density drops off like $r^{-k_\rho}$ away
from it.  Formally this can only be consistent with our approximations
of a spherical, $\rho\propto r^{-k_\rho}$ cloud if every HII region is
born at the cloud center.  We consider this to be merely a
technicality.  While our global cloud model refers only to a mean
density profile and a turbulent energy, our model for the local
density enhancement reflects the fact that clusters are born at the
peaks of the turbulent density profile -- which will not always occur
at the cloud center.  Indeed, we model mass loss and recoil pressure
as if HII regions peppered the entire cloud and commonly vent out its
side, and we ignore the interaction between regions.

Although our local $k_\rho=1$ density is an idealization, we consider
it an improvement over previous works which assumed homogeneous cloud
gas.  Moreover, the most luminous HII regions, those which dominate
the energy injection, are in reality likely to be born near cloud
centers, and once they
have expanded to a radius that is significantly larger than the
distance from their center to the peak of the density distribution in
the GMC, the approximation that they are expanding down a $k_\rho = 1$
density gradient will be quite good.

\subsubsection{Merging of HII regions} \label{HII-merging}

The shell will continue to expand until it decelerates to the point
where the expansion velocity is comparable to the current velocity
dispersion in the GMC, at which point the shell will break up, merge
with the general turbulent motions, and cease to be a coherent entity.
The HII region experiences a velocity dispersion $\sigma(r) = \sigma
(r/\Rc)^{-k_\sigma}$ after expanding to radius $r$, and we approximate
that the merger occurs when $\sigma(r_{\rm sh})=\dot{r}_{\rm sh}$. If
the cloud properties were to remain fixed during the expansion, the
merger radius of an HII region would be
\begin{equation} \label{Rmerge-HII}
r_m= \left(\frac{2
  p_{\rm sh}}{\Mc\sigma}\right)^{1/(3-k_\sigma-\krho)\rightarrow 2/5} \Rc,  
\end{equation}
applying $\krho\rightarrow 1$ and $k_\sigma\rightarrow-1/2$.  The
factor of 2 arises from our idealization of the cloud as a sphere and
the HII region as a hemisphere.  Equation (\ref{Rmerge-HII}) holds
whether merging occurs in the driving phase ($p_{\rm sh}\rightarrow
p_{\rm sh}(t)$) or the snowplow phase ($p_{\rm sh}\rightarrow p_{\rm
sh}(t_{\rm ms})$). However, since the cloud velocity dispersion and
radius can change during the expansion of an HII region, equation
(\ref{Rmerge-HII}) holds only approximately.

Note that gravity has little effect on HII regions except
during merging, because their shells travel faster than $\sigma(r)$ up
to that point.  Indeed, the crossing time $r/\sigma(r)$ is $2.0
\alpha(r)^{-1/2}$ times the free-fall time on scale $r$.  Since
$\alpha(r)\geq 1$, gravity only marginally affects shell motions. For
this reason, we are justified in continuing to use our similarity
solution (which neglects gravity) up to the point at which a shell merges.

\subsubsection{Energy injection} 
\label{eninjection}

At any given time, the cloud energy terms $\calt$ and $\calm$ include
both the turbulent energy from merged HII regions, and the kinetic
energy of those that are still expanding.  During its expansion,
the kinetic energy of a single HII region is
\begin{equation} \label{calT_II}
\calt_1=\frac12 \psh \drsh + 2\pi \PII \rsh^3,   
\end{equation} 
the two terms reflecting bulk and thermal energy, respectively, if
$\PII$ is the internal pressure.  The bulk kinetic energy at the time
of merging $t_m$ is therefore $\calt_1(t_m) =\psh\sigma(r_m)/2$.

Consider first the energy injected into turbulence at merging.  We
assume a merging shell stores the same fraction of energy in
non-kinetic form as does the background turbulence, so we consider the
total energy contribution to be $1.6\calt_1(t_m)$.    Rather than
track individual regions' contributions in detail, we lump them
together in the energy budget in a way that reproduces the correct
average energy in turbulence.  Since the cloud's energy dissipation
time is $t_{\rm dis} = 2.0 (1.2\phi_{\rm in}/\etav) \Rc/\scl$, whereas
the dissipation of a single contribution occurs over $t_{\rm
  dis,1}=2.0(1.2\phi_{\rm in}/\etav)r_m/\sigma(r_m)$, we add 
\begin{equation} \label{deltaE_implemented}
\Delta E_{\rm turb,1} = \etae \frac{r_m \scl}{\Rc \sigma(r_m)} \times 1.6
\calt_1\rightarrow  
\etae \left(\frac{r_m}{\Rc}\right)^{1/2} \times1.6 \calt_1, 
\end{equation} 
rather than $1.6 \calt_1$, to the turbulent energy. The factor $\etae$ parameterizes our ignorance of the exact efficiency with which HII regions drive motions in molecular clouds. For a uniform medium $\etae = 1$, but there has been some suggestion that $\etae < 1$ in turbulent media \citep[e.g.][]{elmegreen00, dale05}, although other simulations do not appear support this result \citep[e.g.][]{mellema05}. In our models we vary $\etae$ to see explore how sensitively our results depend on the efficiency with which HII regions inject energy.

We exclude from consideration the energy that expanding shells possess
prior to merging (which we estimate to be several times smaller than
the mean turbulent energy contributed by these regions).  We note that
the cloud outside of a given region is unaffected by it, except
gravitationally, until it has merged. 
Thus, the rate of energy injection due to a single HII region is roughly
\begin{equation}
\label{HII-injected}
\calg_{\rm cl} = 1.6 \etae \calt_1 \left(\frac{r_m}{\Rc}\right)^{1/2}
\delta(t-t_m),
\end{equation}
and when a shell merges with the overall cloud 
turbulence 
we set 
\begin{equation}
\label{sigmaneweqn}
\frac{3}{2}\sigma_{\rm new}^2 = \frac{3}{2}\sigma_{\rm old}^2+
\etae \frac{\calt_1}{\Mc}
\left(\frac{r_m}{ \Rc}\right)^{1/2}. 
\end{equation}

\subsubsection{Cometary HII regions and cloud disruption}\label{cometHII} 

Finally, note that it is possible that a shell will contain enough
momentum to expand to the point where its radius is larger than the
cloud radius. If the expansion velocity of the shell at this point is
larger than the escape velocity of the cloud (i.e. $\dot{r}_{\rm
sh}>\sqrt{2 G \Mc/\Rc}$), then the shell will simply unbind the
cloud. This criterion for cloud disruption is somewhat different from
those adopted by \citet{williams97} and \citet{matzner02}; we discuss
the distinction and its implications in \S~\ref{fiducialresult}.
If the expansion velocity is smaller than the escape velocity,
the shell will deform the cloud into a cometary
configuration \citep{bertoldi90}, but what happens at that point
depends on whether or not the shell is still being driven by an active
association. An undriven shell will neither gain nor lose energy once
$r_{\rm sh} > \Rc$, and its energy will eventually be added to the
cloud's as turbulence once the cloud falls back. We therefore
approximate that any shell in the snowplow phase that reaches $r_{\rm
sh} > \Rc$ but does not have enough momentum to unbind the GMC merges
with the turbulence. On the other hand, a shell with
an active source can continue to gain energy and evaporate mass even
after reaching the cometary configuration, although eventually its
energy will saturate. Following \citet{williams97}, we
estimate that a shell can continue gaining energy after reaching the
cometary phase up to a time $t \sim 2 t_R$, where $t_R$ is the time it
took the HII region to reach $\Rc$. Its radius at this time will be
$r_{\rm sh} = 1.74 \Rc$. If a driven HII region reaches this radius
and still does not have enough energy to unbind the cloud, then it can
affect the cloud no further. We approximate that the HII region merges
with the turbulence and evaporates no further mass after that point.

It is important to point out that our treatment of large HII regions,
and our criteria for cloud disruption and deformation into a cometary
configuration, are quite approximate. In comparison to
\citet{matzner02}, the criteria we use here reduce the shell momentum
required to disrupt a cloud by roughly a factor of 2. This may affect
our conclusions about the frequency of disruption by HII regions.

\subsection{Supernovae and Protostellar Outflows}

Here we discuss in more detail two sources of feedback that we have chosen not to include: supernovae and protostellar outflows. We omit these because they deliver much less energy per unit mass of stars than HII regions. Here we summarize the arguments, many of which are given in \citet{matzner02}, as to why this is the case. 

The dynamics of a supernovae exploding inside HII regions, and the
amount of energy and momentum they add, may be studied directly either
by numerical simulation \citep[e.g.][]{tenoriotagle85,yorke89} or
analytic calculation \citep{matzner02}. Both methods yield similar
results: supernovae generally sweep up enough mass inside their
ionized bubbles to become radiative by the time they reach the outer
boundary of their host HII regions. As a result, supernova blast waves
radiate away much of their energy before encountering any molecular
cloud material, and consequently increase the total mass removed by
their HII region by at most $\sim 20-40\%$, and more typically $\sim
10\%$. Moreover, this applies only to supernova progenitors whose HII
regions are enclosed by the cloud and do not blister. Supernovae inside
blister-type regions will deposit most of their energy outside the
cloud, into the low density ambient medium, and should therefore have
an even smaller effect. The assumption sometimes made in the literature
\citep[e.g.][]{clark05} that supernovae can halt star formation and
disrupt molecular clouds in $\ltsim 10$ Myr does not appear to be
supported by any numerical or analytic calculations.

Protostellar outflows are negligible compared to HII regions on the size scale of entire GMCs for two reasons. First, the momentum injected into a cloud by an outflow is of order $\dot{M}_* v_w$, where $\dot{M}_*$ is the star formation rate and $v_w \sim 40$ km s$^{-1}$ is the rough momentum input per unit mass of stars formed for hydromagnetic winds \citep{matzner00, richer00}. In contrast, as the analysis of \S~\ref{hiiregions} shows, the momentum carried by HII regions is of order $\dot{M}_{\rm phot} c_{\rm II}$, where $\dot{M}_{\rm phot}$ is the rate at which ionizing photons photoevaporate cloud mass. While $v_w$ is larger than $c_{\rm II}$ by roughly a factor of 4, the low star formation efficiencies of GMCs imply that the rate at which ionizing photons evaporate gas must exceed the rate at which the gas transforms into stars by a factor of $\gtsim 10$, so $\dot{M}_{\rm phot} \gg \dot{M}_*$. Our results in \S~\ref{results} confirm this conclusion. Thus, HII regions provide much more momentum than outflows.

Second, outflows generally inject energy on significantly smaller scales than HII regions. While there are some spectacular examples of HH objects parsecs in size, these appear to be exceptions. \citet{quillen05} find in NGC1333 that the typical size of protostellar outflow cavities is only $\sim 0.1-0.2$ pc. This is a low- to intermediate-mass star forming region, and cavities are likely to be even smaller in the denser environments where most Galactic star formation takes place. This is much smaller than the sizes comparable to the cloud radius reached by large HII regions \citep{matzner02}. Following the discussion in \S~\ref{turbdecay}, we expect the energy injected by these outflows to decay much more quickly than the large-scale motions created by HII regions, and therefore make an even smaller contribution that comparison of the momenta would suggest.

It is worth noting that both arguments show that HII regions dominate over outflows only apply if we are concerned with objects comparable in size to entire GMCs. For smaller objects like the gas clumps forming individual clusters, the time for a single HII region to expand may be comparable to the evolution time of the entire region, and much of the energy of the HII region may be deposited outside the clump. As a result, outflows may well dominate for these objects \citep[e.g.][]{quillen05, li06b}.

\section{Star Formation and HII Region Creation}
\label{starformation}

To know how HII region feedback affects GMC evolution, we must know
the star formation rate and the clustering properties of the stars
formed. To estimate the former, we use the turbulence-regulated star
formation model of
\citet{krumholz05c}. This model postulates that stars form in any
sub-region of the cloud where the local gravitational potential energy
exceeds the turbulent kinetic energy, and it gives good agreement with
simulations, with the observed star formation rate in the Milky Way
\citep{luna06}, with the age spreads of rich star clusters \citep{tan06a},
with the star formation rate in dense molecular clumps \citep{krumholz06c},
and with the the Kennicutt-Schmidt Law for star formation in galactic
disks \citep{kennicutt98b}. In the Krumholz \& McKee model, the star
formation rate is
\begin{equation}
\label{sfreqn}
\dot{M}_* = \sfrff \frac{\Mc}{\tff},
\end{equation}
where
\begin{equation}
\sfrff \approx 0.073 \avir^{-0.68}
\calm^{-0.32},
\end{equation}
$\tff\equiv[3\pi/(32 G \bar{\rho})]^{1/2}$ is the free-fall time at
the mean density $\bar{\rho}$ of the cloud, and $\avir$ and $\calm$ are
the virial parameter and 1D Mach number. Thus, given the mass,
radius, velocity dispersion, and gas temperature in a GMC, the
Krumholz \& McKee model enables us to predict the instantaneous rate
of star formation in that cloud.

HII regions will be driven by the combined ionizing luminosity of all
the stars in an association, and the energy injection therefore
depends on how the stars are
clustered. Thus we must estimate the ionizing luminosity
function of associations as well as the star formation rate. This
distribution is unfortunately quite uncertain, because a small cloud
cannot form arbitrarily large associations and thus the
Galactic-average and individual-cloud luminosity functions are
different. Let $dF_a(\nstar)/d\ln \nstar$ be the fraction of OB
associations with $\ln$ number of stars between $\ln \nstar$ and $\ln
\nstar + d\ln\nstar$ in the entire Galaxy, and let
$dF_{a,\Mc}(\nstar)/d\ln \nstar$ be the corresponding fraction within
a cloud of mass $\Mc$. \citet{mckee97} show that observations of the
Galactic population of HII regions are consistent with a Galactic
distribution
\begin{equation}
\label{galacticfstar}
\frac{dF_a}{d\ln \nstar} \propto \frac{1}{\nstar}.
\end{equation}
It will be more convenient for us to work with the mass of an
association rather than the number of stars. Since we are concerned
with OB associations which give rise to HII regions, and these have
enough stars to sample the IMF well
\citep[$\nstar>100$,][]{zinnecker93}, we convert from number to mass
simply by multiplying by the mean stellar mass. We use the
\citet{kroupa01a} IMF, which gives $\left\langle m\right\rangle_{\rm
IMF}=0.21$ $\msun$, where $\left\langle\right\rangle_{\rm
IMF}$ indicates an average over the IMF. Thus,
\begin{equation}
\label{galacticfma}
\frac{dF_a}{d\ln M_a} \propto \frac{1}{M_a}
\end{equation}
is the Galactic distribution of association mass $M_a$.

To go from the Galactic-average distribution to the
individual cloud distribution, we follow \citet{williams97} and
\citet{matzner02}. First, we require that (1) no association has a
mass larger than $\epsilon \Mc$, where \citet{williams97} estimate
$\epsilon=0.1$, (2) the distribution of association masses within
individual GMCs, when integrated over the GMC population of the
Galaxy, gives the Galactic distribution (\ref{galacticfma}), and (3)
the star formation rate per cloud scales with the cloud mass as
$\dot{M}_* \propto \Mc^{\beta}$. Using an argument analogous to the
derivation of equation (16) in \citet{williams97} and equation (35) in 
\citet{matzner02}, we derive a distribution
\begin{equation}
\label{massocdist}
\frac{d F_{a,\Mc}}{d\ln M_a} \propto
\frac{H(\epsilon \Mc-M_a)}{1 - (\epsilon \Mc/M_a)^{\alpha-\beta}}
\left(\frac{1}{M_a}\right).
\end{equation}
Here $H(x)=(1,0)$ for $(x>0,x<0)$ is the Heaviside step function, and
the Galactic population of star-forming GMCs satisfies
$d\mathcal{N}_{\rm cl}/d\ln \Mc \propto \Mc^{-\alpha}$ with
$\alpha\approx 0.6$ to an upper limit of
$M_u=6\times 10^6$ $\msun$ \citep{williams97}. 
Based on their star formation model,
\citet{krumholz05c} suggest $\beta\approx 0.67$, and we adopt this
value in our work. Note that $\beta<1$ implies a higher rate of star
formation per unit mass in smaller clouds, which occurs because
in the Galaxy smaller clouds have higher densities and thus shorter
free-fall times. Equation (\ref{massocdist}) is fairly simple to
understand intuitively: small clouds cannot make arbitrarily
large associations, hence the factor $H(\epsilon \Mc-M_a)$. If all
clouds produced OB associations with the same mass distribution as the
Galactic powerlaw distribution, up to the maximum mass they could
produce, there would be a deficit of large
associations because all clouds can make small associations, but only
a small fraction of clouds can make large ones. To offset this effect
and produce the Galactic distribution of OB association masses, GMCs
must be slightly more likely to produce an association near their
upper mass limit than a straightforward extrapolation of the Galactic
association mass distribution would suggest, hence the factor
$[1-(\epsilon \Mc/M_a)^{\alpha-\beta}]^{-1}$.

For large associations, the ionizing luminosity is simply
\begin{equation}
S_{49} =
\frac{\left\langle s_{49}(m)\right\rangle_{\rm IMF}}
{\left\langle m\right\rangle_{\rm IMF}} M_a,
\end{equation}
and the main sequence ionizing lifetime is 
\begin{equation}
\left\langle t_{\rm ms} \right\rangle_{\rm a} =
\frac{
\left\langle s_{49}(m) t_{\rm ms}(m) \right\rangle_{\rm IMF}}
{\left\langle s_{49}(m)\right\rangle_{\rm IMF}},
\end{equation}
where $s_{49}(m)$ and $t_{\rm ms}(m)$ are the ionizing luminosity (in
units of $10^{49}$ photons s$^{-1}$) and main-sequence lifetime of a
star of mass $m$. We adopt the fits of \citet{parravano03} for $s(m)$
and $t_{\rm ms}(m)$, which, together with the \citet{kroupa01a} IMF,
give $\left\langle s_{49}(m)\right\rangle_{\rm IMF} = 7.2\times
10^{-4}$, $S_{49} = 3.4\times 10^{-3} (M_a/\msun)$,
and $t_{\rm ms} = 3.8$ Myr. However, for smaller
associations the ionizing luminosity is likely to be dominated by the
single most massive star, and this causes the ionizing luminosity
function to flatten and the ionizing lifetime to vary with
$s_{49}$. \citet[Appendix A]{mckee97} give an analytic formula that
approximates the flattening, but for semi-analytic models we can
dispense with the approximation and determine the luminosity function
simply by drawing stars randomly from the IMF until we accumulate
mass up to a given association mass. We can then determine the
ionizing luminosity of the association by summing those of the individual
stars, and define the main sequence lifetime as the time at which the
stars providing half the ionizing photons disappear. 

Note that, in contrast to \citet{williams97} and \citet{matzner02}, we
do not impose an absolute upper limit of $S_{49} \leq 490$ on the
ionizing luminosity of associations. This has no effect at all on any
but the most massive clouds, since the requirement that the
association mass be at most 10\% of the cloud mass imposes a limit on
$S_{49}$ that is lower than this. For the most massive clouds we
model in \S~\ref{fiducialresult}, the largest associations formed reach
$S_{49} \sim 1700$, but the extremely weak dependence of HII region
properties on $S_{49}$ ($r_{\rm sh}\propto S_{49}^{1/5}$) means that
increasing the maximum $S_{49}$ by a factor of a few has very little
effect. Furthermore, due to the relative improbability of forming an
association so close to the upper mass limit, even for our highest
mass models the majority of clouds do not form associations with
$S_{49} > 490$. Thus, we do not expect the presence or absence of an
upper limit on association ionizing luminosities to affect our results
in any significant way.

\section{Semi-Analytic Models}
\label{method}

\subsection{Methodology}

We now have in place all the necessary theoretical apparatus to set up
our semi-analytic models. In essence, our model describes the
evolution of a GMC using a pair of coupled non-linear ODEs, with added
damping and driving terms.  We integrate these equations forward in
time in a three-step process. First, we use the current configuration of
the cloud to compute the rate of turbulent decay (equation
\ref{Lambda_diss}), the rate of star formation (equation
\ref{sfreqn}), and the rate of evaporation by HII regions (equation
\ref{Mev}). We update $\Rc$, $\dRc$, $\scl$, and $\Mc$ using these
values (equations \ref{eqofmotionfinal} and \ref{energyeqfinal}). We
chose our update time step so that no quantity changes by more than
0.1\% per advance. Second, we update the state of HII
regions as described in \S~\ref{hiiregions}. We compute new values of the
radius and expansion rate for each HII region, removing those whose
expansion rates are low enough or radii are large enough for them to
merge. At merging, we add their energy to turbulent motions in the
cloud.

Third, we create new HII regions. To do this we generate a random
mass for the next association, chosen from the distribution
(\ref{massocdist}). We track the amount of mass transformed into
stars since the last assocation was fully formed, and when half the
next association mass has been accumulated in new stars, we add an HII
region for that association. To determine the properties of the HII
region, we generate stellar masses from the \citet{kroupa01a} IMF,
assign an ionizing luminosity and main sequence lifetime to each star,
and use these to compute the total ionizing luminosity and lifetime
(defined as the time when the stars responsible for half the ionizing
photons burn out) for the association. We continue to put newly formed stars
into the new association until the full mass of the association has
been accumulated, at which point we re-set the tally of accumulated
stellar mass and randomly generate a new mass for the next
association.

We terminate the evolution when one of three conditions has been
satisfied: (1) an HII region unbinds the cloud
(i.e. $r_{\rm sh} \geq R$ and $\dot{r}_{\rm
sh} > \sqrt{2 G M/R}$); (2) the cloud surface density has dropped to
the point where its visual extinction $A_V < 1.4$, the minimum
required for CO to remain molecular \citep{vandishoeck88}, assuming
the standard Milky Way interstellar UV field and dust to gas ratio,
whereby 1 g cm$^{-2}$ corresponds to $A_V = 214$; or (3) the time step
is less than $10^{-8}$ times the current evolution time, which
occurs if the radius is approaching zero. We term these
possibilities \textit{disruption}, \textit{dissociation}, and
\textit{collapse}. An important caveat that applies to all these
outcomes is their dependence on our assumption of spherical
symmetry. Even if an HII region delivers an impulse capable of
unbinding a cloud, the cloud may actually be displaced as a whole, or it may 
break into multiple pieces, each of
which is internally bound and capable of continuing to form
stars. In the case of dissociation, a GMC's mean column density may be
so low that much of its mass is turned atomic by the interstellar UV
field, but overdense clumps within it may survive and continue star
formation. Finally, the collapse case would probably result in a cloud
fragmenting into smaller pieces rather than undergoing monolithic
collapse as occurs in our one-dimensional models.

\subsection{Fiducial Initial Conditions}

Here we describe a basic set of initial conditions for our runs. In
\S~\ref{results}, we discuss how varying some of these affects the
outcome of our runs. Our models start with a common set of initial
parameters summarized in Table \ref{parameters}. Most of these values
are taken directly from observations, but a few of the parameters
deserve some additional discussion. Observations of the strength of
magnetic fields in GMCs are quite uncertain. We set $\etab=0.5$,
corresponding to equipartition between magnetic and kinetic energy.
For a more detailed analysis of the observational data and theoretical
arguments for this choice, see \citet[ \S~7.3]{krumholz05c}.  We
set $\etav=1.2$ and $\phi_{\rm in}=1.0$, corresponding to the decay of
turbulence at the rate found by the simulations of \citet{stone98},
and to turbulence on the size scale of the entire GMC. Finally, we set $\etae=1$, corresponding to HII regions injecting energy into turbulent media as efficiently as they would for smooth media.

\begin{deluxetable}{cc}
\tablecaption{Fiducial parameters.\label{parameters}}.
\tablewidth{0pt}
\tablehead{
\colhead{Parameter} &
\colhead{Value}
}
\startdata
$\azero$ & 1.1 \\
$\cs$ & 0.19 km s$^{-1}$ \\
$\etab$ & 0.5 \\
$\etae$ & 1.0 \\
$\etav$ & 1.2 \\
$\krho$ & 1.0 \\
$\phi_{\rm in}$ & 1.0
\enddata
\end{deluxetable}

With these parameters fixed, we can fully specify the initial
conditions for a model by giving the initial mass and column density
of a cloud. We evolve models with masses of $M_{\rm cl,6}=0.2$, $1.0$ and
$5.0$, where $M_{\rm cl}=M_{\rm cl,6} \times 10^6$
$\msun$. These masses span the range where most of the molecular gas in the
Milky Way resides \citep{williams97}. We set the initial cloud column
density to $N_{H,22} = 1.5$, where $N_H = N_{H,22} \times 10^{22}$
cm$^{-2}$, which is typical for Milky Way GMCs regardless of mass
\citep{larson81, solomon87}. We summarize the initial radius, velocity
dispersion, and crossing time as a function of mass in Table
\ref{cmlist}. For reference, we also show how these quantities vary
with column density at fixed mass.

\begin{deluxetable}{ccccc}
\tablecaption{Initial cloud properties.\label{cmlist}}
\tablewidth{0pt}
\tablehead{
\colhead{$M_{\rm cl,6}$} &
\colhead{$N_{\rm H,22}$} &
\colhead{$\Rczero$ (pc)} &
\colhead{$\sclzero$ (km s$^{-1}$)} &
\colhead{$\tcrzero$ (Myr)}
}
\startdata
0.2 & 0.5 & 33.7 & 2.37 & 13.9 \\
1.0 & 0.5 & 75.3 & 3,54 & 20.8 \\
5.0 & 0.5 & 168  & 5.30 & 31.1 \\
0.2 & 1.5 & 19.4 & 3.12 & 6.1 \\
1.0 & 1.5 & 43.5 & 4.66 & 9.1 \\
5.0 & 1.5 & 97.2 & 6.97 & 13.6 \\
0.2 & 4.5 & 11.2 & 4.10 & 2.7 \\
1.0 & 4.5 & 25.1 & 6.14 & 4.0 \\
5.0 & 4.5 & 56.1 & 9.18 & 6.0
\enddata
\end{deluxetable}

\section{Results}
\label{results}

\subsection{Fiducial Runs}
\label{fiducialresult}

We simulate the fiducial initial conditions 100 times each, using
different random seeds for each run, for each of our three initial
masses. Table \ref{fiducialoutcome} summarizes the
statistical outcome of our fiducial runs.

\begin{deluxetable*}{ccccccccc}
\tablecaption{Fiducial run outcomes.\label{fiducialoutcome}}
\tablewidth{0pt}
\tablehead{
\colhead{$M_{\rm cl,6}$} &
\colhead{$t_{\rm life}$ (Myr)} &
\colhead{$\overline{\alpha}_{\rm vir}$} &
\colhead{$\overline{N}_{H,22}$} &
\colhead{SFE} &
\colhead{$M_{\rm phot}$} &
\colhead{$N_{\rm disrupt}$} &
\colhead{$N_{\rm dissoc}$} &
\colhead{$N_{\rm col}$}
}
\startdata
0.2 & 1.6 (9.9) & 2.2 & 1.4 & 0.053 & 0.59 & 63 & 37 & 0 \\
1.0 & 2.2 (20)  & 2.1 & 1.3 & 0.054 & 0.70 & 92 & 8  & 0 \\
5.0 & 3.2 (43)  & 1.5 & 1.5 & 0.082 & 0.80 & 99 & 1  & 0
\enddata
\tablecomments{
Col. (2): Mean lifetime in crossing times (Myr). Cols. (3-4):
$\avir$ and $N_{H,22}$ averaged over all times and runs. Col. (5): Star
formation efficiency. Col. (6): Fraction of mass photoevaporated prior to cloud destruction.
Cols. (7-9): Number of runs out of 100 that
ended in disruption, dissociation, and collapse.
}
\end{deluxetable*}

\begin{figure}
\plotone{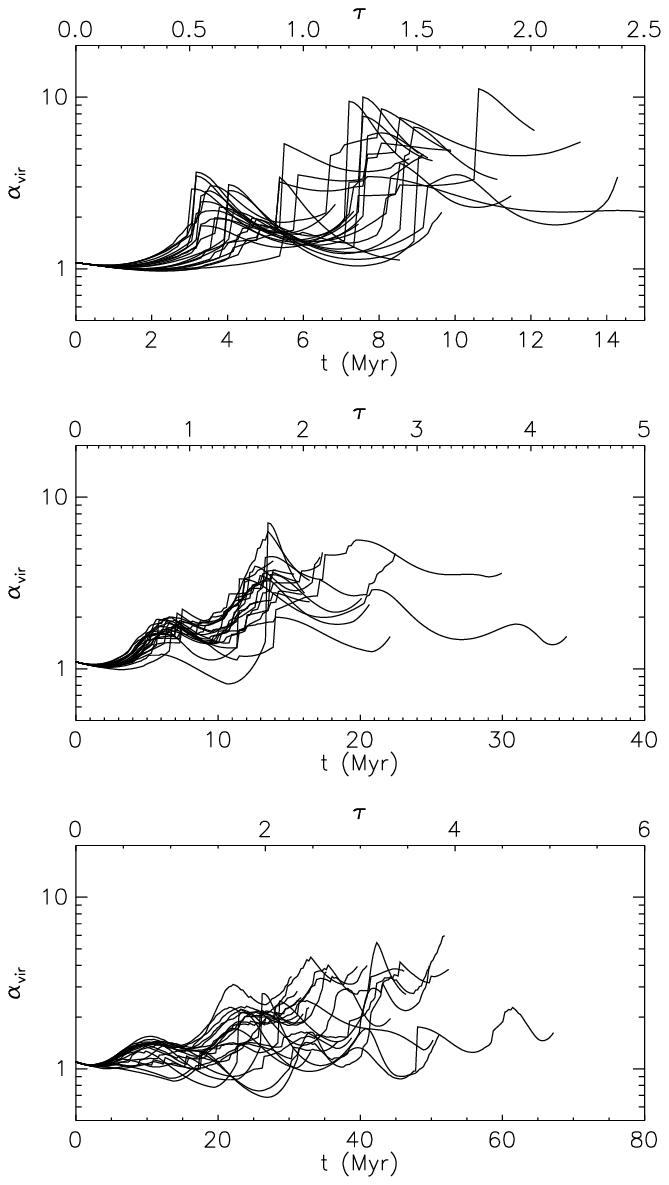}
\caption{
\label{alphaevol}
Virial parameter $\avir$ versus physical time $t$ and dimensionless
time $\tau$ for a sample of runs with fiducial parameters. We show
$M_{\rm cl,6}=0.2$ (\textit{top panel}), $M_{\rm cl,6}=1.0$
(\textit{center panel}), and $M_{\rm cl,6}=5.0$ (\textit{bottom
panel}).
}
\end{figure}

The most basic result is that massive clouds attain a
quasi-equilibrium state, in which the decay of turbulence is roughly
balanced by the injection of energy by HII regions. In this state,
cloud virial parameters fluctuate around unity, but most clouds spend
most of their lives with virial parameters from $1$ to $3$, with a
time-averaged value from $1.5-2.2$ depending on the cloud mass, as
shown by Figure \ref{alphaevol}. This may slightly overestimate the
true virial parameter, because in our model we assume that all of the
energy from HII regions goes into random turbulent motions that can
then fuel cloud expansion, rather than into coherent motions of the cloud as a whole or, if HII regions fragment the cloud, into motions of these fragments. Such coherent motions are often excluded in
observational estimates of virial parameters. Nonetheless, at a
qualitative level the result that
clouds equilibrate to virial parameters $\sim 1$ is in good agreement
with observations showing that massive clouds are approximately
virialized \citep{heyer01,blitz06a}.

\begin{figure}
\plotone{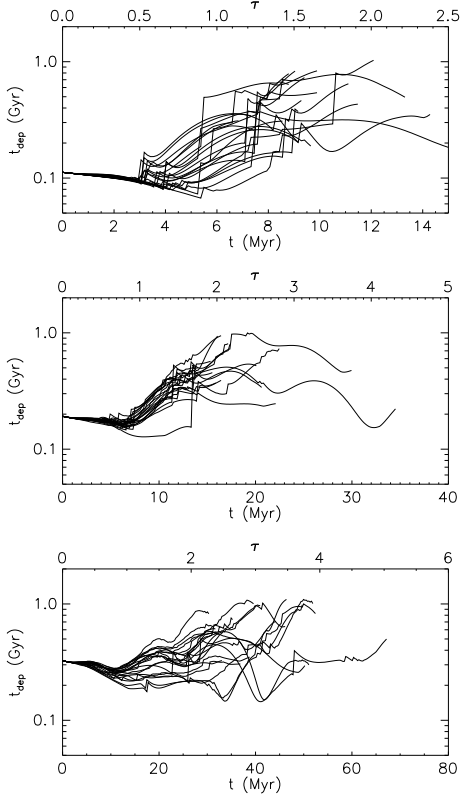}
\caption{
\label{tdepevol}
Depletion time versus physical time $t$ and dimensionless
time $\tau$ for a sample of runs with fiducial parameters. We show
$M_{\rm cl,6}=0.2$ (\textit{top panel}), $M_{\rm cl,6}=1.0$
(\textit{center panel}), and $M_{\rm cl,6}=5.0$ (\textit{bottom
panel}).
}
\end{figure}

The evolution follows a sawtooth
pattern, in which an HII region drives up the virial parameter, which then
exponentially decays until it is increased by the next injection of
energy. This is similar to the pattern seen in simulations by
\citet{li06b}, in which turbulence in star-forming clumps $\sim 1$ pc
in size is maintained by
energy injection from protostellar outflows. This equilibrium is
maintained partly because the star formation rate responds to the
current state of the cloud, increasing as the cloud contracts and its
turbulence decays, and going back down when the cloud
re-expands. Figure \ref{tdepevol} shows the depletion time, defined as
the ratio of the current cloud mass to the current star formation
rate, which exhibits this pattern.

\begin{figure}
\plotone{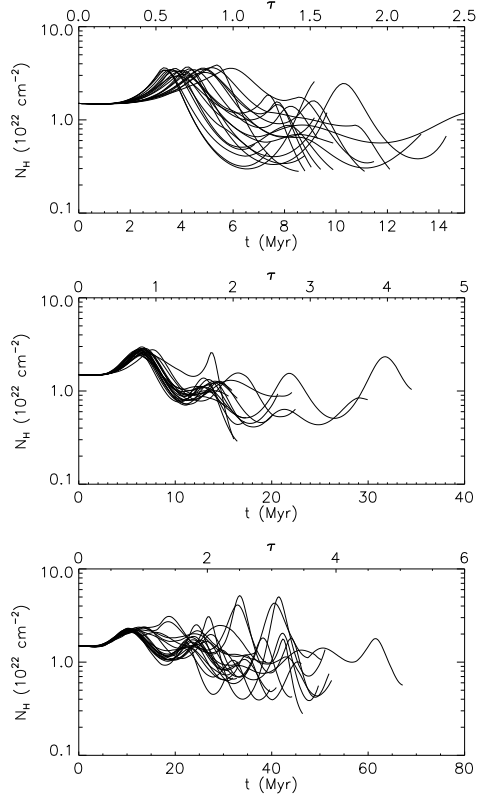}
\caption{
\label{nhevol}
Column density $N_H$ versus physical time $t$ and dimensionless
time $\tau$ for a sample of runs with fiducial parameters. We show
$M_{\rm cl,6}=0.2$ (\textit{top panel}), $M_{\rm cl,6}=1.0$
(\textit{center panel}), and $M_{\rm cl,6}=5.0$ (\textit{bottom
panel}).
}
\end{figure}

\begin{figure}
\plotone{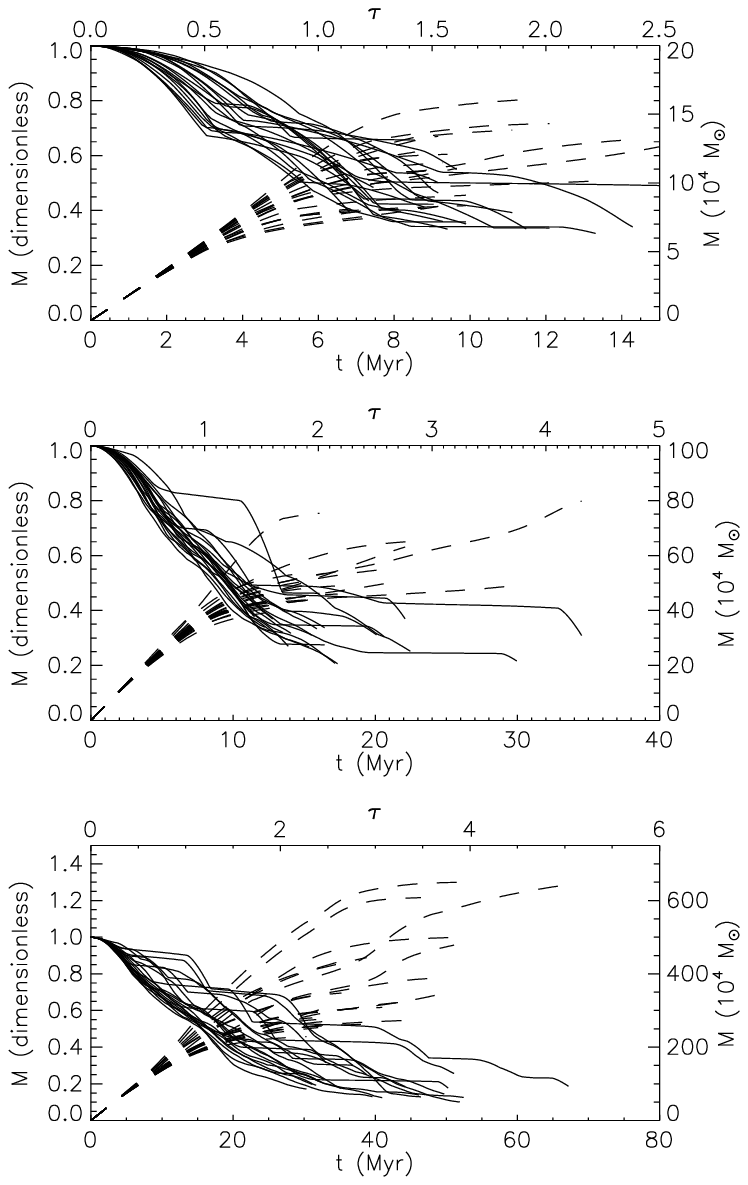}
\caption{
\label{massevol}
Cloud mass (\textit{solid lines}) and 10 times mass of stars formed
(\textit{dashed lines}) versus physical time $t$ and dimensionless
time $\tau$ for a sample of runs with fiducial parameters. We show
$M_{\rm cl,6}=0.2$ (\textit{top panel}), $M_{\rm cl,6}=1.0$
(\textit{center panel}), and $M_{\rm cl,6}=5.0$ (\textit{bottom
panel}).
}
\end{figure}

As shown in Figure \ref{nhevol},
cloud column densities also show a sawtooth pattern, oscillating
up and down but remaining relatively constant over multiple expansion
and contraction cycles so that the time average is roughly the
observed value $N_{H,22}\approx 1.5$. (The figure is somewhat deceptive, because the longest lived-clouds tend to go to somewhat lower column densities, and these stand out the most when examining the figure because the shorter-lived clouds all overlie one another on the left side of the plot. The numerically-computed time-averaged density over all models is given in Table \ref{fiducialoutcome}.)
Over their lifetimes, the
average star formation efficiency of clouds, defined as the fraction
of initial cloud mass transformed into stars, is $5-10\%$. HII regions
ionize away anywhere from $50-90\%$ of the mass before finally
destroying the clouds entirely, with lower mass clouds losing less
mass to photoionization than more massive clouds. Figure
\ref{massevol} shows this evolution.

The duration and stability of cloud equilibrium is affected by
cloud mass. Low mass clouds, $M_{\rm cl,6}=0.2$ are only stable
for an average of $1.6$ crossing times ($3.2$ free-fall times), and
are most often destroyed when they form an HII region that delivers
enough momentum to completely unbind them. They survive only one or a
few cycles of expansion and contraction. In contrast, very massive
clouds, $M_{\rm cl,6}=5$, are stable for $3.2$ crossing times
and for multiple cycles before finally being unbound by HII
regions. At all masses most clouds are destroyed by direct disruption
rather than by a photodissociation, with the number photodissociated
dropping sharply as a function of mass. This result is largely a
function of cloud lifetimes. Direct disruption by a single HII region
occurs when a truly large association forms, one that is on the tail
of the mass distribution. Since larger clouds live longer and form
more stars, they sample this tail more thoroughly, and thus are more
likely to form one very large association capable of disrupting them
than smaller clouds. An additional effect comes from the cloud
velocity dispersion. Since larger clouds have higher velocity
dispersions, the HII region shells driven by small associations merge
more quickly, so the relative amount of energy injected by large
versus small associations increases.

Cloud disruption is much more frequent in our models than in those of
\citet{williams97} or of \citet{matzner02}.  The primary difference is a
revised criterion for dynamical disruption. Williams \& McKee took the
onset of a cometary phase to be the point at which an HII region
effectively disrupts or displaces its parent cloud.  Matzner
considered the delivery of a momentum $p_{\rm sh}$ in excess of $\Mc
v_{\rm esc}$ to define disruption.  Our criterion ($\dot{r}_{\rm
sh}>v_{\rm esc}$
when $r>\Rc$) resembles the Matzner threshold, but requires roughly
half as much momentum because we model HII regions as hemispheres and
clouds as spheres.  In addition, our adoption of the \citet{kroupa02}
IMF and \citet{parravano03} stellar properties leads to about twice as many
ionizing photons per stellar mass relative to Williams \& McKee and
Matzner. The revised model of HII region expansion also plays a minor
role.  All told, the largest cloud that can be disrupted by HII
regions is about ten times more massive in this work than in the
Matzner models.  Our disruption criterion, being more generous than
the Williams \& McKee or Matzner criteria, accounts in an approximate
manner for cloud displacement and fragmentation by large HII regions.

Our conclusion that more massive GMCs survive longer than less massive
ones also contrasts with the results obtained \citet{williams97} and
\citet{matzner02}. One reason for our different
result is that in our model the star formation rate per unit mass is
proportional to the inverse of the free-fall time, which, at constant
$N_H$, produces more star formation per unit mass in small clouds than
in large ones. Thus, low mass clouds are subject to more vigorous
feedback. In contrast, previous studies generally adopted star formation
laws under which the star formation rate per unit mass was independent
of or increased with GMC size. This higher star formation rate
compensates for the lack of large associations in small clouds and
the ease with which small clouds can be driven into a cometary
configuration, both of which tend to reduce mass loss. As a result
clouds photoevaporate about the same fraction of their mass per
dynamical time regardless of their starting mass.

A second reason for our finding that small clouds live less time is
our focus on the dynamical rather than the photevaporation lifetime.
For massive clouds, our models show that the
photoevaporation $e$-folding time is $\sim 20-30$ Myr, comparable to
or a bit shorter than the dynamical lifetime, and consistent with
earlier estimates. In contrast, for low mass clouds the dynamical
destruction time is shorter than the photoevaporation time. Clouds are
dynamically broken up by HII regions, either by direct disruption from
a single region or expansion to the point of photodissociation through
the collective action of several HII regions, before most of their
mass is photoevaporated away. This occurs because smaller clouds have
less inertia and lower escape velocities, so they are more
easily pushed apart by HII regions. Larger clouds require truly
large HII regions to unbind them, and so are not disrupted until they
have lost significant mass and enough associations have formed to
reach to this tail of the distribution. Smaller clouds can be
destroyed by more typical HII regions.

This distinction between photoevaporation and dynamical lifetime
points to an important caveat of our analysis. The dynamical lifetime
we find is the time for which a cloud exists as a single coherent
entity under the assumption of spherical symmetry. As we discuss
above, it is possible that, rather than
unbinding all the molecular gas, or expanding it to the point of
photodissociation, HII regions displace GMCs or blow GMCs into multiple
pieces that are not bound to each other, but each of which remains
molecular and internally bound, and can continue forming stars. Thus,
the time for
which the Lagrangian elements of a cloud remain molecular and
star-forming may be longer than the dynamical lifetime of the cloud
which we have found. The photoevaporation time may provide
a better estimate of this Lagrangian lifetime. For clouds of mass
$M_{\rm cl,6}\gtsim 1$ this distinction is not that significant since
the lifetime is comparable to the photoevaporation time, but
it can be for smaller clouds. Determining whether
parts of these clouds do continue forming stars even after they no
longer form a single gravitationally bound entity will require
radiation hydrodynamic simulations that can include both cloud
dynamics and photoevaporation. 

\subsection{Varying Column Density}
\label{varcolsection}

Prompted in part by the result that giant molecular clouds both in the
Milky Way \citep{larson81,solomon87} and in nearby galaxies with
similar metallicities \citep{blitz06a} have a
surprisingly small range of column densities, we investigate how
varying the initial column density affects the evolution. For each
mass we run with the fiducial parameters, we re-run using the same 100
random seeds but initial column densities of $N_{H,22} = 0.5$ and $4.5$.
The low column density case corresponds to a GMC with a mean column
density that is just barely sufficient to be self-shielding against
the interstellar UV field. Giant clouds of such low column density (as
opposed to isolated low-mass clouds) have not been seen in CO surveys
either in the Milky Way or in other local group galaxies, and there is
observational evidence and theoretical expectation that molecular gas
at column densities lower than $N_{N,22}\sim 0.9$ is not star-forming
(\citealt{mckee89, li97, onishi98, onishi99, onishi02, johnstone04} --
though see \citealt{hatchell05}, who find less star formation at low column densities, but not a complete absence of star formation). However, in the model of GMCs as turbulent
density fluctuations, low column density clouds of all masses are
expected to exist and be star-forming
\citep[e.g.][]{vazquezsemadeni97}, so we consider the low column
density case without imposing a column density threshold for star
formation. The high column density case corresponds to GMCs found in
galaxies that are entering the starburst regime. For example, in M64,
a weak nuclear starburst galaxy, the typical GMC column density is 2.5
times that in the Milky Way \citep{rosolowsky05a}.

\begin{figure}
\plotone{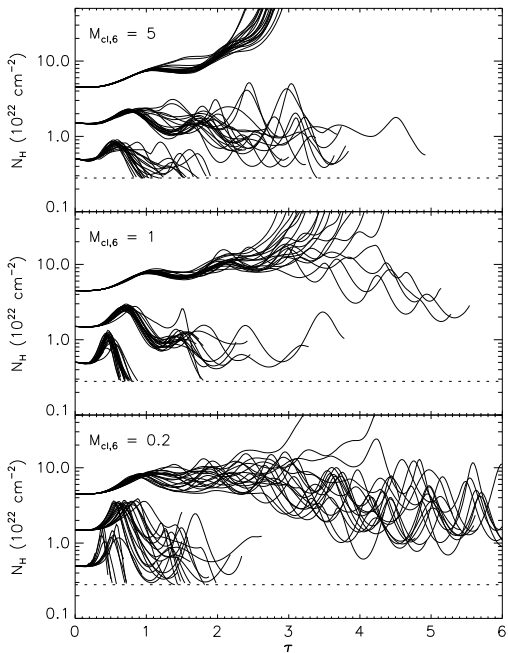}
\caption{
\label{nhvary}
Column density $N_H$ versus dimensionless time $\tau$ for a sample of
runs with varying initial $N_H$. Each panel shows a different initial
mass, shown in the upper left corner. The dotted horizontal line
indicates the column density at which clouds dissociate.
}
\end{figure}

Table \ref{nhvarytab} gives statistical results for the
runs with varying $N_H$, and Figure \ref{nhvary} shows
the evolution of column density versus time for a sample of runs at
each mass. The results show that $N_{H,22}\approx 1$ seems to
be roughly a critical point in column density. Clouds that begin their
evolution at substantially lower column density tend to disrupt or
dissociate in $\sim 1$ dynamical time. Clouds that start at higher
column densities show a pattern that depends on mass. At masses
$M_{\rm cl,6} \ll 1$, they remain stable for long times. At higher
masses, they tend to undergo uncontrolled collapse.

\begin{deluxetable*}{cccccccccc}
\tablecaption{Outcomes with varying column density.\label{nhvarytab}}
\tablewidth{0pt}
\tablehead{
\colhead{$M_{\rm cl,6}$} &
\colhead{$N_{\rm N,22}$} &
\colhead{$t_{\rm life}$ (Myr)} &
\colhead{$\overline{\alpha}_{\rm vir}$} &
\colhead{$\overline{N}_{H,22}$} &
\colhead{SFE} &
\colhead{$M_{\rm phot}$} &
\colhead{$N_{\rm disrupt}$} &
\colhead{$N_{\rm dissoc}$} &
\colhead{$N_{\rm col}$}
}
\startdata
0.2 & 0.5 & 0.44 (6.1) & 1.7 & 0.60 & 0.022 & 0.34 & 74 & 26 & 0 \\
0.2 & 1.5 & 1.6 (9.9) & 2.2 & 1.4 & 0.053 & 0.53 & 63 & 37 & 0 \\
0.2 & 4.5 & 5.6 (15) & 1.5 & 5.2 & 0.19 & 0.80 & 93 & 0 & 7 \\
1.0 & 0.5 & 0.72 (15) & 1.9 & 0.61 & 0.026 & 0.51 & 8 & 92 & 0 \\
1.0 & 1.5 & 2.2 (20)  & 2.1 & 1.3 & 0.054 & 0.70 & 92 & 8  & 0 \\
1.0 & 4.5 & - & - & - & - & - & 17 & 0  & 83 \\
5.0 & 0.5 & 1.3 (41)  & 1.6 & 0.5 & 0.039 & 0.58 & 5 & 95  & 0 \\
5.0 & 1.5 & 3.2 (43)  & 1.5 & 1.5 & 0.082 & 0.80 & 99 & 1  & 0 \\
5.0 & 4.5 & - & - & - & - & - & 0 & 0 & 100
\enddata
\tablecomments{
Column definitions are identical to those in Table
\ref{fiducialoutcome}, and cases with $N_{H,22}=1.5$ are identical to
the values in that Table. We compute average quantities excluding
runs that result in collapse, and we do not attempt to compute
averages in cases where a majority of runs produce collapse.
}
\end{deluxetable*}

This change in behavior is not due to a change in the star formation
rate per dynamical time. This is almost independent
of the column density, since $\sfrff \propto \calm^{-0.32} \propto
N_H^{-0.08}$ for fixed $\avir$. Instead, the evolution depends on the
column density because column density significantly affects the
efficiency of feedback. With our approximation that merged HII regions do not inject energy into clouds, the fraction of an HII
region's energy that is lost to radiation increases with column
density, so for equal ionizing fluxes and times, the kinetic energy in
an HII region shell varies as $N_H^{-3/5}$. Furthermore, higher column
densities
increase the cloud velocity dispersion (as $N_H^{1/4}$ for fixed
$\avir$) and decrease the expansion velocity, causing HII regions to
break up and merge earlier. Since in our model HII regions only gain energy
as long as they are expanding, the sooner they merge the less energy
they inject. From equation (\ref{Rmerge-HII}), for fixed cloud and HII
region properties the ratio of the merger radius to the cloud radius
scales as $N_H^{-9/50}$. Combining these two effects, equation
(\ref{HII-injected}) shows that the energy injected by a single HII
region of fixed properties into a cloud of fixed mass varies with
column density as $N_H^{-69/100}$. Thus, the efficiency with which
ionizing photons are converted into turbulent motions is a reasonably
strong function of the column density, and this picks out a particular
characteristic column density at which energy injection balances
loss. This turns out to be roughly the observed column density
$N_{H,22} \approx 1.5$. At this point we should mention one
significant cautionary note: we have not explored the effects of the
external environment of GMCs, and in particular the ambient pressure,
on the characteristic column density. We will consider this effect in Paper II.

It is easy to understand intuitively why the critical column density
above which clouds tend to collapse for masses $M_{\rm cl,6}
\gtsim 1$ should be between $N_{H,22} = 1.5$ and $4.5$. In the high
column density case, the velocity dispersion required to hold up a
cloud is almost more than HII regions can provide. HII regions cannot
effectively drive turbulence to velocity dispersions larger than the
ionized gas sound speed $\cii=9.7$ km s$^{-1}$, and as indicated in
Table \ref{cmlist}, in the high mass, high column density case the
level of turbulence required to maintain $\avir=1.1$ is
$\sclzero=9.2$ km s$^{-1}$. For the medium mass case it is
$\sclzero=7.0$ km s$^{-1}$. Furthermore, if the cloud contracts some
so that its velocity dispersion increases a bit beyond these values,
then HII regions will break up and merge with the turbulence almost
immediately and will therefore inject very little
energy. \citet{matzner02} previously discussed the inability of HII
regions to maintain virial balance in systems where the required
velocity dispersion exceeds $\cii$, and our models find the same
result. We discuss the implications of this in more detail in
\S~\ref{gmcevol}. Note, however, that this conclusion is partially
dependent on our approximation that HII regions cease driving
turbulence once they cease dynamically expanding and merge, which may
not be entirely correct -- see \S~\ref{limitations}.

The long lifetime and high star formation efficiency
produced at low masses and high column density also suggests an
interesting interpretation: this combination of parameters may
correspond to the regime of formation of individual OB associations
and open clusters, which is characterized by relatively high column
densities \citep{mckee03} and star formation efficiencies $\gtsim 10\%$
\citep{lada03}, and probably requires many crossing times to complete
\citep{tan06a}. Although the highest column density we have considered
is still relatively modest by the standards of some cluster-forming
clumps, which can reach $N_{H,22}$ of several tens, the general result that
low masses and high column densities can produce long-lived bound
objects is suggestive. This is particularly true in light of the
recent evidence that rich clusters require $\gtsim 5$ crossing times
to assemble \citep{tan06a, huff06, krumholz06c}, and must therefore
be held up against collapse yet not be unbound by feedback for at
least this long.

\subsection{Varying Dissipation Rate}
\label{dissvarysec}

As we discuss in \S~\ref{turbdecay}, it is possible that the turbulent
dissipation rate may vary from our fiducial estimate, either being
substantially lower if simulations of turbulent decay have missed some
important physics, or substantially higher if the characteristic scale
of molecular cloud turbulence is smaller than the size of an entire
GMC. The latter possibility is probably ruled out by observations
showing that turbulence in molecular clouds is self-similar out to the
size of the entire GMC \citep[e.g.][]{ossenkopf02, heyer04}, but we
explore it nonetheless to better understand how the dissipation rate
affects GMC evolution. We re-run our fiducial case $\etav=1.2$,
$\phiin=1.0$, corresponding to turbulence on the GMC scale and decay at
the rate measured by \citet{stone98}, with $\etav=0.4$,
$\phiin=1.0$, corresponding to turbulence on the GMC scale decaying
somewhat more slowly than Stone et al.\ find, and with
$\etav=1.2$, $\phiin=0.33$, corresponding to turbulence driven on
$1/3$ of the GMC size scale decaying at the Stone et al.\ rate. Since
the dissipation rate depends on the ratio $\etav/\phiin$, we are
therefore considering energy dissipation rates that are a factor of 3
smaller and larger than the fiducial case. (From equations
\ref{eqofmotionfinal}, \ref{energyeqfinal}, and \ref{Lambda_diss},
this factor of 3 larger of smaller dissipation rate should correspond
roughly to a factor of 3 change in the acceleration of the cloud
radius, since $\call\propto \etav \sigma^3$, $\sigma'\propto
\call/\sigma^2$, and $R''\propto \sigma^2$.)

\begin{figure*}
\epsscale{0.85}
\plotone{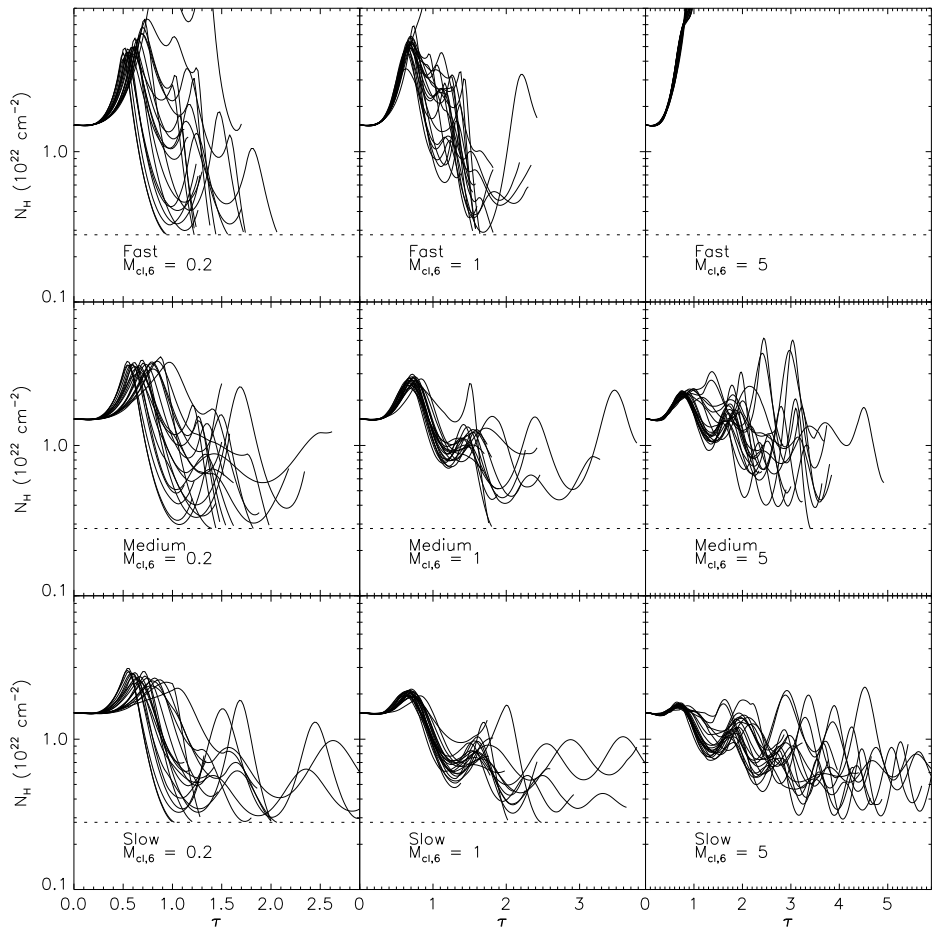}
\epsscale{1.0}
\caption{
\label{dissvaryfig}
Column density $N_H$ versus dimensionless time $\tau$ for a sample of
runs with varying dissipation rates and masses. The mass and
dissipation rate are indicated in each panel -- slow is $\etav=0.4$,
$\phiin=1.0$, medium is $\etav=1.2$, $\phiin=1.0$, and fast is
$\etav=1.2$, $\phiin=0.33$. The dotted horizontal line indicates the
column density at which clouds dissociate. All runs begin with
$N_{H,22} = 1.5$.
}
\end{figure*}

Table \ref{dissvarytab} summarizes the statistical results of varying
the dissipation rate, and Figure \ref{dissvaryfig} shows the evolution
of column density versus time as a function of the dissipation rate
and cloud mass. The general result is that, with the exception of the
high mass, fast dissipation case, the results do not greatly depend on
the dissipation rate. As the turbulent dissipation rate increases, the
lifetime and mean virial parameter decrease and the mean column
density and the star formation efficiency increase, but none of these
changes by more than $\sim 10\%$. That changing the
dissipation rate by a factor of three in either direction induces a
much smaller change in the cloud column density and virial parameter
suggests that these values represent roughly an equilibrium
configuration, and that this equilibrium is quite robust. As the
dissipation rate changes by a factor of nine from the lowest to the
highest values we try, clouds contract a bit more, form stars
a bit more vigorously, and are destroyed by stellar feedback a bit
sooner, but only modest changes are sufficient to offset the change in
dissipation rate. Star formation self-regulates to produce GMCs with
$N_{\rm H,22}\approx 1$ and $\avir\approx 1-2$, and the regulation is
stiff in the sense that even an order of magnitude change in the
dissipation rate does not alter it much. We discuss reasons for this
in \S~\ref{paramdependence}.

\begin{deluxetable*}{ccccccccccc}
\tablecaption{Outcomes with varying dissipation rate.\label{dissvarytab}}
\tablewidth{0pt}
\tablehead{
\colhead{$M_{\rm cl,6}$} &
\colhead{$\etav$} &
\colhead{$\phiin$} &
\colhead{$t_{\rm life}$ (Myr)} &
\colhead{$\overline{\alpha}_{\rm vir}$} &
\colhead{$\overline{N}_{H,22}$} &
\colhead{SFE} &
\colhead{$M_{\rm phot}$} &
\colhead{$N_{\rm disrupt}$} &
\colhead{$N_{\rm dissoc}$} &
\colhead{$N_{\rm col}$}
}
\startdata
0.2 & 0.4 & 1.0  & 1.7 (10)  & 2.8 & 1.2 & 0.046 & 0.53 & 49 & 51 & 0 \\
0.2 & 1.2 & 1.0  & 1.6 (9.9) & 2.2 & 1.4 & 0.053 & 0.59 & 63 & 37 & 0 \\
0.2 & 1.2 & 0.33 & 1.4 (8.4) & 1.9 & 1.8 & 0.067 & 0.57 & 44 & 56 & 0 \\
1.0 & 0.4 & 1.0  & 2.3 (21)  & 2.6 & 1.1 & 0.046 & 0.64 & 90 & 10 & 0 \\
1.0 & 1.2 & 1.0  & 2.2 (20)  & 2.1 & 1.3 & 0.054 & 0.70 & 92 & 8  & 0 \\
1.0 & 1.2 & 0.33 & 1.7 (16)  & 1.7 & 2.1 & 0.074 & 0.74 & 61 & 35 & 4 \\
5.0 & 0.4 & 1.0  & 4.4 (60)  & 2.3 & 1.0 & 0.070 & 0.75 & 98 & 2  & 0 \\
5.0 & 1.2 & 1.0  & 3.2 (43)  & 1.5 & 1.5 & 0.082 & 0.80 & 99 & 1  & 0 \\
5.0 & 1.2 & 0.33 &       -   &   - &   - &     - & - &  0 & 0  & 100
\enddata
\tablecomments{
Column definitions are identical to those in Table
\ref{fiducialoutcome}, and cases with $\etav=1.2$, $\phiin=1.0$ are
identical to the values in that Table. As in Table \ref{nhvarytab}, we
compute average quantities excluding runs that result in collapse, and
we do not attempt to compute averages in cases where a majority of
runs produce collapse.
}
\end{deluxetable*}

There are two exceptions to this, where varying the dissipation
rate does produce a qualitative change in the outcome. For clouds of high
mass and high dissipation rate, we see a phenomenon similar to the
high mass, high column density case. For $M_{\rm cl,6}=5$, the
velocity dispersion starts at $\sclzero = 7.0$ km s$^{-1}$, but
increases very rapidly as the cloud contracts due to decay of
turbulence. By the time the first large HII regions have expanded to
the point where they contain substantial kinetic energy, the cloud has
contracted to the point where the velocity dispersion required to hold
it up is comparable to or larger than $\cii=9.7$ km s$^{-1}$. As a
result, HII regions cannot hold up the cloud, and it collapses.

\begin{figure*}
\epsscale{0.85}
\plotone{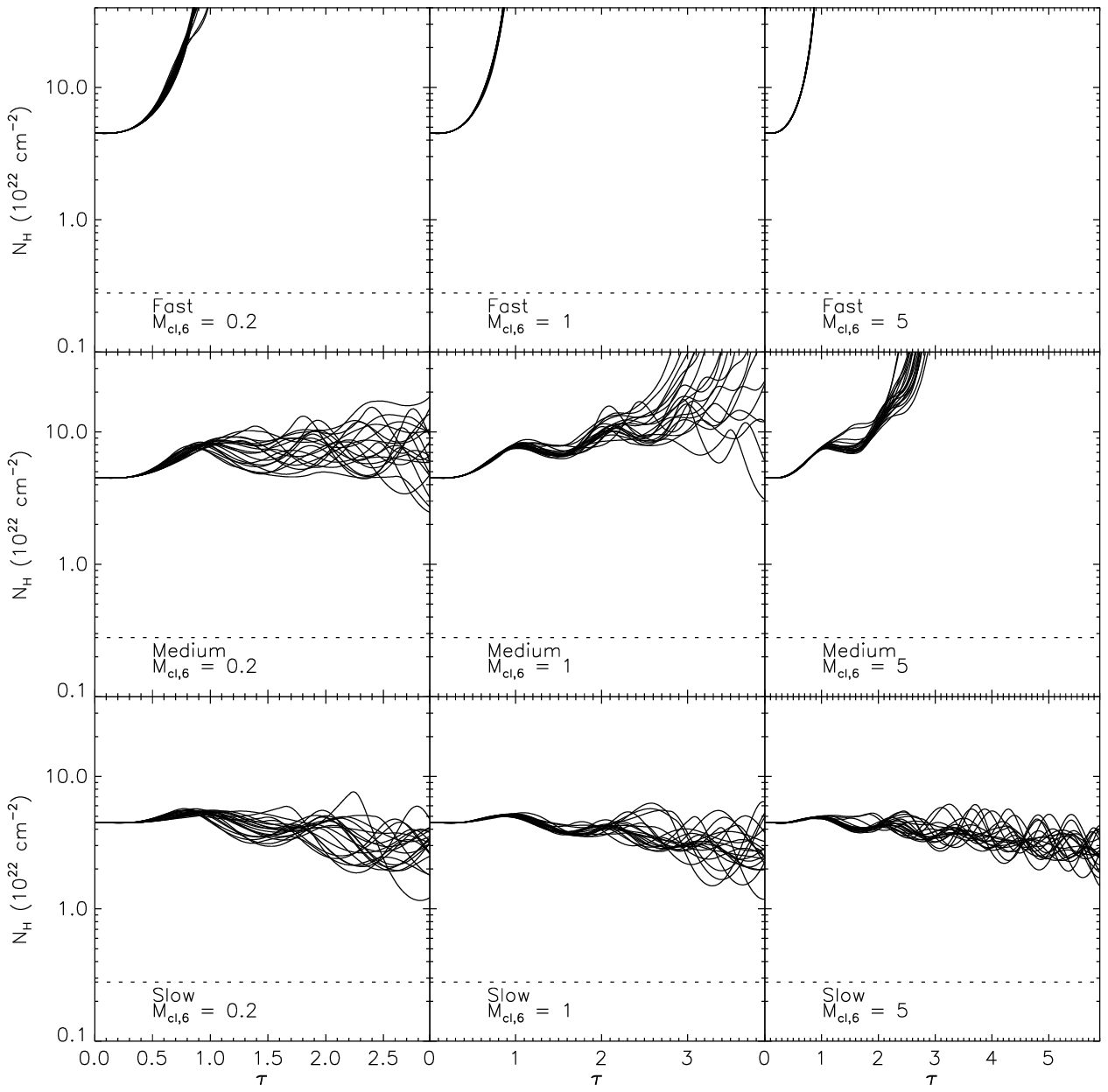}
\epsscale{1.0}
\caption{
\label{nhevol.dissvary.nhhigh}
Column density $N_H$ versus dimensionless time $\tau$ for a sample of
runs with varying dissipation rates and masses, all starting from high initial column densities. The mass and
dissipation rate are indicated in each panel -- slow is $\etav=0.4$,
$\phiin=1.0$, medium is $\etav=1.2$, $\phiin=1.0$, and fast is
$\etav=1.2$, $\phiin=0.33$. The dotted horizontal line indicates the
column density at which clouds dissociate. All runs begin with
$N_{H,22} = 4.5$.
}
\end{figure*}

The other exception occurs if we simultaneously increase the column
density to $N_{H,22}=4.5$ and lower the dissipation rate by setting
$\etav=0.4$, $\phiin=1.0$. In this case, clouds have long lifetimes
and high star formation efficiencies regardless of their
mass, and their column densities gradually decrease with
time. Qualitatively, this is the same behavior we see for the case
$M_{\rm cl,6}=0.2$, $N_{H,22}=4.5$, and the fiducial dissipation
rate. Figure \ref{nhevol.dissvary.nhhigh} shows the evolution of
clouds within initial column densities of $N_{H,22}=4.5$ as a function
of mass and dissipation rate. As the plot shows, reducing the
dissipation rate changes the characteristic mass at which one goes
into the high star formation efficiency, long-lived regime. The
boundary between this regime of very long stability and the regime of
stability for a few dynamical times appears to depend weakly on both
the mass and the dissipation rate.

\subsection{Varying Energy Injection Efficiency}
\label{effvary}

As we discuss in \S~\ref{eninjection}, we are not certain of the
efficiency with which HII regions are able to drive motions in
turbulent media. We therefore re-run our fiducial models with
$\etae=0.25$, thereby reducing the amount of energy injected by HII
regions by a factor of 4. This allows us to examine how strongly our
results depend on our assumed efficiency.

\begin{figure*}
\epsscale{0.85}
\plotone{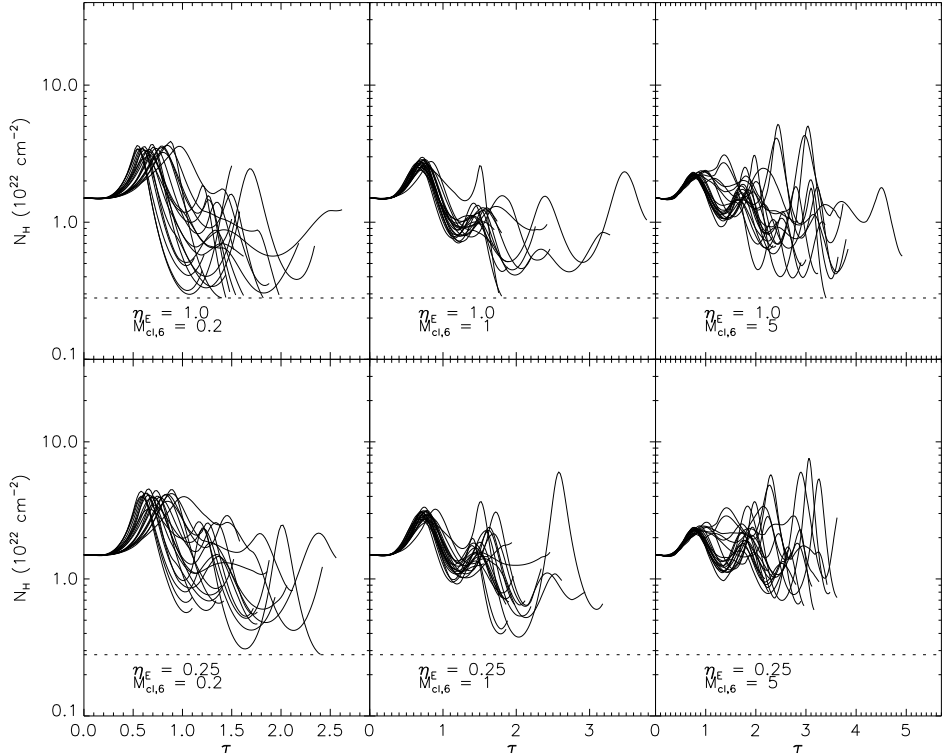}
\epsscale{1.0}
\caption{
\label{nhevol.effvary}
Column density $N_H$ versus dimensionless time $\tau$ for a sample of
runs with varying HII region driving efficiencies and masses. The mass and
driving efficiency are indicated in each panel. The dotted horizontal line indicates the
column density at which clouds dissociate. All runs begin with
$N_{H,22} = 1.5$.
}
\end{figure*}

Table \ref{effvarytab} summarizes our results, which show how small a
difference the change in driving efficiency makes. With a factor of 4
less energy injection for an HII region of the same luminosity, the
mean lifetime of the lower mass clouds we model increases by about
10\% and that of the highest mass clouds decreases by the same
fraction. Mean virial parameters decline by $\sim 10\%$, and mean
column densities rise a similar amount. Both star formation
efficiencies and fractions of the mass photoevaporated rise, but again
by only $\sim 10\%$. The only quantity that changes significantly is
the fraction of clouds destroyed by dissociation, which not
surprisingly declines sharply. Thus, our results appear extremely
insensitive to changes in the assumed energy injection efficiency of
HII regions. Figure \ref{nhevol.effvary}, which shows the evolution of
column density versus time for a sample of our runs, confirms this
impression. The runs with lower energy injection efficiency go to
slightly higher column densities, but overall show no major difference
in their evolution from the fiducial case. Again, the star formation
process seems to be self-regulating, so that large changes in
efficiencies, either of energy injection or of dissipation, produce
only very small changes in the global statistics of cloud
evolution. We discuss reasons for this in \S~\ref{paramdependence}.

\begin{deluxetable*}{cccccccccc}
\tablecaption{Outcomes with varying HII driving efficiency.\label{effvarytab}}
\tablewidth{0pt}
\tablehead{
\colhead{$M_{\rm cl,6}$} &
\colhead{$\etae$} &
\colhead{$t_{\rm life}$ (Myr)} &
\colhead{$\overline{\alpha}_{\rm vir}$} &
\colhead{$\overline{N}_{H,22}$} &
\colhead{SFE} &
\colhead{$M_{\rm phot}$} &
\colhead{$N_{\rm disrupt}$} &
\colhead{$N_{\rm dissoc}$} &
\colhead{$N_{\rm col}$}
}
\startdata
0.2 & 0.25 & 1.7 (11) & 1.6 & 1.7 & 0.065 & 0.67 & 93 & 7 & 0 \\
0.2 & 1.0 & 1.6 (9.9) & 2.2 & 1.4 & 0.053 & 0.59 & 63 & 37 & 0 \\
1.0 & 0.25 & 2.1 (20) & 1.6 & 1.6 & 0.062 & 0.75 & 100 & 0 & 0 \\
1.0 & 1.0  & 2.2 (20)  & 2.1 & 1.3 & 0.054 & 0.70 & 92 & 8  & 0 \\
5.0 & 0.25 & 2.8 (39) & 1.3 & 1.8 & 0.087 & 0.82 & 100 & 0 & 0 \\
5.0 & 1.0  & 3.2 (43)  & 1.5 & 1.5 & 0.082 & 0.80 & 99 & 1  & 0 \\
\enddata
\tablecomments{
Column definitions are identical to those in Table
\ref{fiducialoutcome}, and cases with $\etae=1.0$ are identical to
the values in that Table.
}
\end{deluxetable*}

\section{Discussion}
\label{discussion}

\subsection{The Relationship of Cloud Properties to Rates of Radiative
Loss and Energy Injection}
\label{paramdependence}

One of the most interesting results of our analysis is how little the
globabl statistics of cloud evolution, such as lifetimes and star
formation efficiencies, change in response to large changes in
parameters such as the rate at which turbulence decays via isothermal
shocks or the efficiency with which HII regions transfer their
kinetic energy into turbulent motions.
We can understand intuitively why large variations in either driving
efficiency or dissipation rate cause such minimal changes in the
statitistics of cloud evolution by examining Figure \ref{tdepevol}. As
the Figure illustrates, clouds form most of their stars during their
contraction phases, when they reach high densities, so the rate of
star formation is effectively set by the cadence of expansion and
contraction cycles. This cadence is affected only slightly by the
dissipation rate and the energy injection efficiency, because both of
these properties are coupled to cloud evolution poorly, with large
time delays.

For a GMC to contract from an expanded state, the
rate-limiting step is not the time required to dissipate the energy
injected by HII regions, it is the cloud free-fall time, the time the
cloud would take to collapse even if it contained no turbulence at
maximum expansion. Thus, varying the turbulent dissipation rate makes
little difference to the cloud contraction time.
Similarly, the time it takes for HII regions to drive apart a cloud is
controlled less by the amount of energy added per unit mass of stars
formed than by the time delay between when stars begin forming and
when the HII regions they create expand, break up, and drive turbulent
motions. Lowering the driving efficiency means that clouds expand somewhat
less far, but since the dissipation rate of the turblence is a strong
function of the velocity dispersion, $\call\propto \sigma^3$, the extra
energy at early stages produced by higher efficiency is radiated away
very quickly, and produces only slightly increased cloud expansion. We
can put this argument more formally by noting that our energy equation
gives $\sigma'\propto \sigma^2$, neglecting changes in velocity dispersion
due to external pressure and cloud expansion, so the time required for
a velocity dispersion of $\sigma_0$ immediately after a large HII
region breaks up to decay to a value $\sigma_1$ obeys $t\propto
(\sigma_0-\sigma_1) / (\sigma_0 \sigma_1)$. If the final velocity
dispersion is much smaller than the initial one, i.e. $\sigma_1 \ll
\sigma_0$, this ratio is independent of $\sigma_0$. Thus, the decay
time for a large amount of initial energy injected is only slightly
larger than the decay time for a significantly smaller energy
injection, so the cloud expansion time and maximum radius depend only
weakly on the driving efficiency.

This insensitivity to the dissipation rate and the energy injection
rate breaks down if the dissipation rate becomes so high or the
energy injection so inefficient that HII regions are not capable of
expanding clouds at all. In this case, clouds undergo runaway
collapse. It also breaks down if the dissipation rate becomes so low
or the efficiency so high that the first generation of HII regions
simply unbinds most clouds, in which case clouds form only one
generation of stars and are destroyed on a time scale set by the
expansion time for HII regions from that first generation. However, in
between these two extremes there is a broad range of parameter space
within which the cadence of expansion and contraction cycles is set
mostly by the cloud free-fall time and the time required for HII
regions to break up, with only a very weak dependence on the details
of the energy loss rate or the energy gain efficiency. The degree of
insensitivity is illustrated by Figures \ref{dissvaryfig} and
\ref{nhevol.effvary}, in which factor of several variations in
dissipation rate or driving efficiency change the expansion and
contraction period by only tens of percent.

\subsection{GMC Stability and Lifetime}
\label{gmcstability}

Observations place strong constraints on the stability of giant
molecular clouds, and are now beginning to constrain their lifetimes
as well. We must test theories that seek to explain the behavior of GMCs
against these observations. The first constraint is that giant
molecular clouds cannot be in a state of global collapse. If they
were, then they would convert order unity of their mass into stars in
a crossing time, and this would produce a star formation rate
two orders of magnitude larger than the observed one
\citep{zuckerman74}. Thus, a model of GMC evolution must explain why
GMCs are stable against global collapse and convert only a few percent of
their mass into stars per crossing time.

The second observational constraint is that GMCs, at least the most
massive ones, live for considerably more than a single crossing
time. While GMC lifetimes are quite difficult to estimate inside the
Milky Way, age spreads of the largest OB associations are $\sim 20$
Myr \citep{blaauw64, blitz80}, which is a probable lower limit on the lifetimes
of GMCs. More robust constraints are available in extragalactic
observations, where there is no distance ambiguity. Associations
between GMCs and star clusters imply a typical GMC lifetime of $\sim
20$ Myr in M33 \citep{engargiola03} and 27 Myr in the LMC
\citep{fukui99,blitz06a}. This is significantly greater than the GMC crossing
time of $\ltsim 10$ Myr. While the difference between one and a few
crossing times might seem insignificant, recall that the $e$-folding
time for the decay of turbulence is roughly a crossing time. A cloud
two crossing times old will have lost almost all of its turbulence if
feedback or some other source cannot replenish it, and will therefore
undergo global collapse. Thus, the observations not only require that
clouds be stable against collapse, they require that this stability be
maintained for several turbulent decay times.

The GMC model we present in this Paper, using the fiducial parameters
suggested by observation and previous theoretical work, provides very
good qualitative agreement with the observations. We show that
feedback from star formation can keep clouds supersonically turbulent
and virialized for $\sim 30$ Myr, until they are destroyed by
feedback. Combined with the results of \citet{krumholz05c} showing
that supersonic turbulent motions naturally produce a star formation
rate of a few percent per crossing time if star formation occurs in
virialized clouds, this model satisfies both observational
constraints. We discuss the question how how GMCs evolve during this
lifetime in more detail in \S~\ref{gmcevol}.

Alternative models run into difficulty with one of these two
observations, or with other data. One possible explanation why GMCs do
not undergo global collapse is that they are gravitationally unbound
transient fluctuations in the atomic ISM; only local subregions
constituting a small fraction of the cloud mass are bound and can
collpase \citep{maclow04, clark04, clark05, vazquezsemadeni06,
dobbs06a}. However, in this case it is hard to see how GMCs could
survive $20-30$ Myr. For example, \citet{clark05} find GMC lifetimes
of only about 10 Myr in their simulations of unbound clouds. There are
also other severe observational difficulties. There are no molecular
clouds with masses $\gtsim 10^4$ $\msun$ that are clearly
gravitationally unbound, with virial parameters $\avir\gg 1$, either
in the Milky Way \citep{heyer04} or in other galaxies
\citep{engargiola03, rosolowsky03, rosolowsky05a, blitz06a}. There are
many examples in the Milky Way of clouds of mass $<10^4$ $\msun$ with
virial parameters $\avir\gg 1$ \citep{heyer04}, and there is no obvious
reason why massive clouds should not also display a range of $\avir$
if they are generally unbound. Transient fluctuation models also have
problems explaining the existence of a GMC mass scale and the low
rotation rates of GMCs. These arguments are presented in detail in
\citet{krumholz05c}.

A second possiblity is that GMCs are bound, at least marginally, but
that they are destroyed before the turbulence with which they are born
decays away. As a result, they never have a chance to begin global
collapse. However, the observed lifetimes of $\gtsim 20$ Myr appear to
rule out the possibility that such rapid destruction is the
norm. Furthermore, to work this model requires a mechanism of cloud
destruction that reliably operates in $\ltsim 1$ crossing time. As we
show here, HII regions are not capable of completely
photoevaporating or disrupting massive clouds over such short
periods. Supernovae and protostellar winds inject consierably less
energy into GMCs per unit mass of stars formed than do HII regions
\citep{tenoriotagle85, yorke89, matzner02}, so they are unlikely
candidates for rapid 
cloud disruption. In the absence of a plausible mechanism for cloud
disruption in $\sim 10$ Myr, this model faces a major theoretical
problem as well as an observational one.

It is worth noting at this point that
arguments for GMC lifetimes in the Milky Way strictly limited to 1
crossing time, e.g. \citet{hartmann01}, generally rely on observations
of clouds in the solar neighborhood, all of which are much smaller than
the larger, more typical GMCs we have considered in this paper. Since
we find that molecular clouds $\ltsim 10^5$ $\msun$ in mass do only
survive for $\sim 1$ crossing time, our results are consistent with
the idea that solar neighborhood GMCs are short-lived. We simply
suggest that GMC lifetime is mass-dependent, and that the nearest
clouds, due to their atypical masses, also have atypical lifetimes.

\subsection{An Evolutionary Scenario for GMCs}
\label{gmcevol}

Our findings allow us to present a rough evolutionary scenario for
GMCs that is consistent both with the constraints of stability and GMC
age described in \S~\ref{gmcstability} and with the observation that
GMCs in the Milky Way and in other galaxies all lie in a narrow
range of column densities and virial parameters, centering around
$N_{H,22} \approx 1.5$ and $\avir \approx
1-2$. \citep{larson81,solomon87, blitz06a}. This scenario is quite
similar to that suggested in \citet{mckee89} and is developed in
considerably greater detail in Paper II.

Our scenario is that GMCs are born at low column densities by
condensation out of the atomic ISM, primarily triggered by
self-gravitational instabilities in spiral shocks
\citep[e.g.][]{kim01, kim02, kim03}. This can only occur in regions of
a galactic disk that are at significantly higher densities and
pressures than is typical of the ISM. If star formation were able to
start in such clouds immediately, they would undergo rapid disruption,
converting $\ltsim 3\%$ of their mass into stars. However, in reality
such clouds are probably not star-forming over most of their volume,
so feedback is suppressed.

This may explain the observation that GMCs
pass through a phase in which they appear to lack embedded HII
regions. The duration of this phase is difficult to obtain from
observations -- \citet{fukui99}, \citet{engargiola03} and
\citet{blitz06a} report that $\sim 1/4$ of GMCs do not show
H$\alpha$ signatures associated with HII regions, but this is probably
an overestimate because it does not include highly obscured HII
regions. Although the H$\alpha$ sensitivities of the catalogs used by
these authors are sufficient to detect HII regions with H$\alpha$
luminosities comparable to the Orion Nebula, Orion is visible largely
because it is on the near side of a dark cloud; at the distance of the
LMC or M33, Orion would be invisible if it were oriented with the dark
cloud on the near side rather than the other way around. Spitzer/MIPS
observations have failed to find any GMCs that are dark at 24 $\mu$m,
so there must be some embedded star formation even in GMCs without
detected HII regions
(E. Rosolowsky, 2006, private communication). This result suggests
either that the HII regions surveys are incomplete, or that clouds do
not begin making massive stars until well after they have begun
forming low mass stars, an effect we have not considered.

Regardless of how long this initial starless phase lasts, a question
we will address using our models in Paper II, in the
absence of feedback, GMCs will contract due to loss of turbulent
support, gravity, and external pressure, until they approach
$N_{H,22}\approx 1$. At
that point, they become star-forming, and feedback stabilizes them
against further contraction and keeps them in virial equilibrium for several
crossing times. We observe GMCs at $N_{H,22}\approx 1$,
$\avir\approx 1-2$ because this is where they spend the vast majority
of their lifetimes. This column density is selected because it is the
one for which energy injection by turbulence roughly balances energy
loss by isothermal shocks. In addition to turbulent energy injection,
the recoil momentum produced by mass being evaporated by HII regions
may also play a significant role in confining clouds.

This quasi-equilibrium endures for an amount of time that depends on
the cloud mass. For massive clouds, which contain most of the
molecular mass in the Milky Way and in all but one other galaxy in
which it is possible to estimate cloud mass distributions
\citep{fukui99,engargiola03,blitz06a}, it endures $2-3$ crossing
times, or $20-30$ Myr. During a cloud's lifetime, it converts $5-10\%$
of its mass into stars. These results are robust against changes in
the assumed dissipation rate for turbulence or efficiency of turbulent
driving by HII regions.
Less massive clouds are probably dynamically disrupted in $\sim 1$
crossing time, consistent with estimates of the lifetimes of small
clouds in the solar neighborhood \citep[e.g.][]{hartmann01},
although parts of them may endure in molecular phase and continue
forming stars for longer periods of time even after the original cloud
has been disrupted.

One interesting question is how our results might change if we
considered a galaxy quite different from the Milky
Way. Observations indicate that GMCs in normal spiral galaxies like
the Milky Way are generally quite similar to those in the Milky Way
\citep{blitz06a}, so to find truly different conditions we must
consider galaxies that are approaching the regime of starbursts. The sole
observational example of such a galaxy where we have observed the
clouds is M64, a weak starburst in which the largest GMCs are $\gtsim
10^7$ $\msun$ in mass, and have surface densities up to $N_H\approx
4\times 10^{22}$ cm$^{-2}$ \citep{rosolowsky05a}. Despite these
differences from Milky Way GMCs, observations indicate that these
clouds are still in approximate virial balance, and that like Milky
Way GMCs their star formation rate is only a few percent per free-fall
time. Thus, they are not in a state of global collapse. 

Since we have found that HII
regions cannot sustain the turbulence and prevent collapse in such
clouds (though see \S~\ref{limitations}), we are left with two
possibilities. Either the dissipation rate in such clouds is lower
than our fiducial estimate, or there is an additional source of energy
input that supplements HII regions. One strong candidate for a source
of energy injection is driving of turbulence by external shocks
\citep{kornreich00}, either from supernovae, gravitational
instabilities driven by the potential of the stars, or cloud-cloud collisions \citep{tan00}. This mechanism
encounters considerably difficulty in the Milky Way because GMCs are
much denser than the gas around them in which shocks propogate. This
creates an ``impedance mismatch'' that makes it difficult to drive
turbulence into the GMCs \citep{nakamura06}. However, in a starburst
where the ISM is
entirely molecular, clouds are not much denser than their
surroundings. \citet{rosolowsky05a} find that in M64 the GMCs are
overdense only by factors $\sim 2$. This removes the impedance
mismatch problem, and makes it far easier for external shocks to drive
turbulence than it is in galaxies like Milky Way. Whether this
mechansim can work in detail will require more analytic models and
simulations to determine.

\subsection{Limitations of This Model}
\label{limitations}

The most obvious limitation of our model is the constraint that GMCs
evolve homologously. In mathematical terms, this constraint is
equivalent to dropping time derivatives of $\ci$ in the virial
theorem. This should only change our results substantially if changes
in the moment of inertia of a GMC occured primarily through changes in
its shape rather than overall expansion or contraction. While a cloud
in approximate equilibrium probably does experience most changes in
its inertia via changes in shape, a cloud that is undergoing global
collapse or global disruption would almost certainly experience most
changes in its inertia through those processes rather than through
changes in shape. As a result, our broad conclusion that neither overall
collapse or disruption occur for several crossing times seems likely
to be robust.

A more subtle limitation to our model is our simple boundary
conditions for GMCs. For simplicity we have assumed a fixed mass
budget and a fixed external pressure that puts GMCs into pressure
balance initially. In reality, GMCs may start forming stars while
still accumulating gas. The somewhat higher mean mass for GMCs with
associated HII regions than for GMCs without such associations seen in
the LMC \citep{fukui99,blitz06a} seems to point in this direction. In
addition, GMCs form in spiral shocks. As a result, the Lagrangian
mass elements from which a cloud forms are probably subjected to a
rising pressure as they enter the arm region, and then experience a
falling pressure as the cloud moves out of the arm. This may have
important effects on GMC evolution, and we explore the effects of more
complex boundary conditions in Paper II. A related issue in GMC
evolution, which we will also explore, is the effect of the apparent
existence of column density thresholds for star formation. If clouds
cannot begin forming stars until they accumulate a certain column of
molecular mass, then feedback in GMCs will not start up until some
time after the cloud first becomes molecular. This lag may affect GMC
evolution.

There are also several limitations associated with our treatment of HII regions. We assume that the mass of the cloud is large enough that its evolution is dominated by HII regions, not protostellar outflows.
We assume that the clouds are spherical, which means that all the energy injected by HII regions goes into internal motions, whereas in fact some of it goes into moving the entire cloud. Finally, we neglect possible energy injection by HII regions after merging. In our model, shells driven by small HII
regions or those in very turbulent clouds may drop to expansion
velocities below the cloud velocity dispersion well before the driving
stars burn out. At this point the HII region will cease to drive a
coherent expanding shell, and we approximate that the energy
injection ceases. However, if the driving stars continue to ionize a
part of the cloud, there is another potential driving
mechanism. Clumps of gas that enter the ionized region will be
photoevaporated on one side, and therefore will be rocketed away from
the ionizing stars. Even though the resulting thrust is not enough to
drive the clumps out of the HII region entirely, since the expansion
velocity is smaller than the cloud velocity dispersion, it can
still alter the trajectories of gas clumps and potentially inject
energy \citep{tan01}. This
effect could be important in the cases where we find that HII regions
cannot support GMCs because the velocity dispersion required to hold
up the cloud is comparable to or larger than the ionized gas sound
speed. Determining whether this effect is significant will require
either more detailed theoretical treatment or numerical
simulations. Despite these caveats, though, our results in
\S~\ref{dissvarysec}-\ref{effvary} and our analysis in
\S~\ref{paramdependence} show that our conclusions regarding the
lifetime and evolutionary history of GMCs are quite robust against
changes in the assumed efficiency with which HII regions drive
turbulent motions or the rate at which those turbulent motions decay.

\section{Conclusions}
\label{conclusions}

In this Paper we derive the basic equations governing the evolution of
the mass, radius, and velocity dispersion of giant molecular clouds
under the approximation of homologous motion. We construct simple
models for the rate at which turbulent motions decay due to radiation
from isothermal shocks and for the rate at which HII regions driven by
massive stars within the cloud drive turbulent motions, and we use these
to study the global evolution and energetic balance of
GMCs. This enables us to build GMC models in which we neither assume
energetic and virial balance, nor neglect the effects of
feedback. Thus, we are able for the first time to address critical
questions such as whether GMCs are in virial or energetic equilibrium,
how long they live, what determines their lifetimes and star formation
histories, and what determines properties such as column density and
virial parameter that appear to be roughly the same for all GMCs.

Our primary conclusion is that giant molecular clouds with observed
properties are indeed quasi-equilibrium objects. The rate at which
they lose energy via radiation from isothermal shocks is roughly
balanced by the rate at which HII regions from star formation inject
energy back into the cloud. This feedback keeps them virialized,
$\avir = 1-2$, and keeps them at the observed column density
$N_H\approx 1.5\times 10^{22}$ cm$^{-2}$. Our results suggest that
this column density, which appears to be generic for GMCs both in the
Milky Way and in other galaxies where GMCs are observable, is in fact
the result of feedback, since the efficiency of star formation
feedback varies with column density, and the observed column density
corresponds to the one required for equilibrium. Whether the GMC
formation process, particularly the passage through an overdense and
overpressured spiral shock, also plays a role in selecting the GMC
column density is at this point an open question, one we will address
in Paper II.

The duration of the equilibrium state of a GMC depends on cloud
mass, but for clouds $\gtsim 10^6$ $\msun$ in mass it lasts for $2-3$
crossing times, or roughly 30 Myr. The dominant destruction mechanism
for GMCs is dynamical unbinding by the momentum delivered by an expanding HII
region, but for massive clouds this unbinding does not happen until
HII regions have photoionized away $\sim 90\%$ of the GMC
mass. Smaller clouds are dynamically disrupted by HII regions in $\sim
10$ Myr while $\sim 50\%$ of their mass is still in molecular form,
but it is unclear if this disruption actually destroys their molecular
material or simply breaks it into smaller, mutually unbound
clouds.

Overall, our models produce a picture of GMCs that is quite different
than that suggested by models in which GMCs are gravitationally
unbound or in which feedback from star formation is neglected. Models
that neglect feedback appear incapable of reproducing the observed
lifetimes and properties of the GMC population. We conclude that any
reasonable model of GMC evolution cannot neglect the effects of star
formation feedback, and that even a simple model of feedback such as
the one we use here produces good agreement with the observed
properties of GMCs.

\acknowledgements We thank E. C. Ostriker, J.~P. Ostriker,
E. Rosolowsky, J.~M. Stone, and J.~C. Tan for helpful discussions, and
the anonymous referees for useful comments. Support for this work was
provided by NASA through Hubble Fellowship grant \#HSF-HF-01186
awarded by the Space Telescope Science Institute, which is operated by
the Association of Universities for Research in Astronomy, Inc., for
NASA, under contract NAS 5-26555 (MRK), by NSERC and the Canada
Research Chairs Program (CDM), and by the NSF through grants
AST-0098365 and AST-0606831 (CFM).

\begin{appendix}

\section{Derivation of the Virial Theorem For an Evaporating
Cloud}
\label{virialderivation}

Here we derive the equation of motion for an evaporating homologous
cloud, which is a generalization of the Eulerian Virial Theorem
\citep[EVT,][]{mckee92}. Gas in the cloud is injected into the wind at
a rate $\drho$, so the continuity equation for the cloud is
\begin{equation}
\label{conteqn}
\frac{\partial \rho}{\partial t} = -\nabla\cdot\rho\mathbf{v}+\drho,
\end{equation}
where $\rho$ and $\mathbf{v}$ are understood to be the density and
velocity of cloud and not wind material. The wind satisfies its own
continuity equation $\partial \rho_w/\partial t =
-\nabla\cdot\rho_w\mathbf{v}_w-\drho$,
so that the summed equation $\partial (\rho+\rho_w) =
-\nabla\cdot(\rho\mathbf{v}+\rho_w\mathbf{v}_w)$ is simply the
ordinary continuity equation. The equation of momentum conservation
for the cloud gas is
\begin{equation}
\label{peqn}
\frac{\partial}{\partial t}(\rho \mathbf{v}) =
-\nabla\cdot(\mathbf{\Pi}-\mathbf{T}_{M}) + \rho\mathbf{g}
+\drho(\mathbf{v}+\vkick),
\end{equation}
where $\mathbf{\Pi}$ is the gas pressure tensor, $\mathbf{T}_{M}
\equiv [\mathbf{B}\mathbf{B}-(1/2)B^2\mathbf{I}]/(4\pi)$ is the
Maxwell stress tensor, $\mathbf{B}$ is the magnetic field, and
$\mathbf{g}$ is the gravitational force per unit mass. Intuitively,
the $\drho\mathbf{v}$ term represents the change in cloud momentum due
to mass being transferred into the wind, and $\drho\vkick$ is the
recoil momentum from the ejection. As with the continuity equation,
there is a corresponding momentum equation for the wind.

We now follow \citet{mckee92} in deriving the EVT for a cloud
with a wind. The moment of inertia of the cloud is
\begin{equation}
\Ic \equiv \int_{\Vv} \rho r^2 \,dV,
\end{equation}
where the fixed volume of integration $\Vv$ is chosen to be sufficiently
large that it includes the entire cloud at all times. Taking the time
derivative
\begin{eqnarray}
\dot{I}_{\rm cl} &=& 
- \int_{\Vv} (\nabla\cdot \rho \mathbf{v}) r^2 \,dV + \int_{\Vv} \drho r^2
\,dV \nonumber\\
&=& - \int_{\Sv} (\rho \mathbf{v} r^2)\cdot d\mathbf{S} + 2 \int_{\Vv}
\rho \mathbf{v}\cdot \mathbf{r} \,dV + \ci \dMc \Rc^2
\end{eqnarray}
where $\Sv$ is the surface of the volume $\Vv$, and
\begin{equation}
\ci \equiv \frac{3-\krho}{5-\krho}.
\end{equation}
Differentiating again,
\begin{eqnarray} \label{IddotE}
\frac{1}{2} \ddot{I}_{\rm cl} &=& - \frac{1}{2} 
\int_{\Sv} r^2 \frac{\partial}{\partial t}(\rho \mathbf{v}) \cdot
d\mathbf{S} +
\int_{\Vv} \frac{\partial}{\partial t}(\rho \mathbf{v}) \cdot
\mathbf{r} \,dV
\nonumber\\
& &
{} + \frac{1}{2} \ci \ddMc \Rc^2 + \ci \dMc \Rc \dRc. 
\end{eqnarray} 

The time derivative can be taken out of the integral in the first term
because $\Sv$ is fixed, and for the second term we can substitute for
the integrand using the momentum equation (\ref{peqn}). This gives
\begin{eqnarray} 
\label{IddotE-2}
\frac{1}{2} \ddot{I}_{\rm cl} 
& = &
-\frac{1}{2} \frac{d}{dt} \int_{\Sv} r^2 \rho \mathbf{v} \cdot
d\mathbf{S}
- \int_{\Vv}
\left\{
\mathbf{r}\cdot
\left[\nabla\cdot(\mathbf{\Pi}-\mathbf{T}_M)\right]
-\rho\mathbf{v}\cdot\mathbf{g}\right\}\,
dV 
\nonumber \\ 
&& {} +
\int_{\Vv} \drho \mathbf{r} \cdot\mathbf{v} \,dV 
+ \int_{\Vv} \drho r v_{\rm ej}' \,dV
+ \frac{1}{2} \ci \ddMc \Rc^2 + \ci \dMc \Rc \dRc . 
\end{eqnarray}
To evaluate the term $\int_{\Vv} \drho \mathbf{r} \cdot\mathbf{v}
\,dV$, we make use of our homology assumption, which allows us
to write the velocity at any point in the cloud as
\begin{equation}
\label{vdecomp}
\mathbf{v} = \mathbf{r} \frac{\dRc}{\Rc} + \mathbf{v}_{\rm turb},
\end{equation}
where the first term represents homologous expansion or contraction of
the cloud, and the second is a turbulent velocity that carries no net
radial flux of matter. Thus, $\int_{\Vv} \rho \mathbf{r}
\cdot\mathbf{v}_{\rm turb} \,dV = 0$, and $\int_{\Vv} \drho \mathbf{r}
\cdot\mathbf{v} = \int_{\Vv} \drho r^2 \dRc/\Rc$.

This allows us to write the final EVT,
\begin{eqnarray}
\frac12 \ddot{I}_{\rm cl} & = & 
2(\calt-\calt_0) + 
\calb+\calw-\frac12
\frac{d}{dt}\int_{\Sv} (\rho\mathbf{v} r^2)\cdot d\mathbf{S} 
\nonumber\\ 
& & {} +
2 \ci \dMc \Rc \dRc + \frac{3-\krho}{4-\krho} \dMc \Rc v_{\rm ej}' +
  \frac12 \ci \ddMc \Rc^2.
\label{EVT}
\end{eqnarray} 
In this equation, the kinetic term is
\begin{equation}
\calt=\frac{1}{2} \int_{\Vv} (3 P + \rho v^2) \,dV
\end{equation}
for gas thermal pressure $P$, the surface term is
\begin{equation}
\calt_0 = \frac{1}{2} \int_{\Sv} \mathbf{r} \cdot
\mathbf{\Pi} \cdot d\mathbf{S},
\end{equation}
the gravitational term is
\begin{equation}
\calw = \int_{\Vv} \rho \mathbf{r}\cdot\mathbf{g} \,dV,
\end{equation}
and the magnetic term is
\begin{equation}
\label{calbdef}
\calb=\frac{1}{8\pi} \int_{\Vv} (B^2-B_0^2) \, dV,
\end{equation}
where $B_0$ is the background magnetic field far from the cloud, and
we require that $\Vv$ be large enough to include the full volume over
which the background field is distorted by the presence of the cloud.

\section{Derivation of Energy Conservation For an Evaporating Cloud}
\label{energyderivation}

Here we derive the equation of energy conservation for an evaporating
homologous cloud. First, we rewrite the momentum equation (\ref{peqn})
in a slightly different form,
\begin{equation}
\label{peqn2}
\rho\frac{d\mathbf{v}}{dt}=-\nabla P - \rho\nabla \phi +
\frac{\mathbf{J}\times\mathbf{B}}{c} +\drho\vkick,
\end{equation}
where the terms on the right hand side are the pressure,
gravitational, and Lorentz forces, $\phi$ is the gravitational
potential, $\mathbf{J}$ is the current density, and we have replaced
the pressure tensor $\mathbf{\Pi}$ with the isotropic pressure
$P\mathbf{I}$ under the assumption that viscosity is
negligible. Taking the dot product of (\ref{peqn2}) with $\mathbf{v}$
yields
\begin{equation}
\label{eneqn1}
\frac{\partial}{\partial t}
\left(\frac{1}{2} \rho v^2\right)
+\nabla\cdot \left(\frac{1}{2} \rho \mathbf{v} v^2\right)
= -\mathbf{v}\cdot\nabla P - \rho\mathbf{v}\cdot\nabla\phi
+\frac{\mathbf{v}}{c} \cdot (\mathbf{J}\times\mathbf{B})
+ \drho \left(\frac{1}{2} v^2 + \mathbf{v}\cdot\vkick\right).
\end{equation}
Using the continuity equation, we can rewrite the gravitational work
term as
\begin{eqnarray}
-\rho \mathbf{v} \cdot \nabla\phi & = &
-\nabla\cdot\rho\mathbf{v}\phi + \phi\nabla\cdot\rho\mathbf{v} \\
& = &
-\nabla\cdot\rho\mathbf{v}\phi - \frac{\partial}{\partial t}(\rho
\phi) + \rho\frac{\partial\phi}{\partial t} + \drho \phi.
\end{eqnarray}
Similarly, we can rewrite the magnetic work term using Poynting's
Theorem, which in the MHD approximation is
\begin{equation}
\frac{1}{8 \pi} \frac{\partial}{\partial t} (B^2-B_0^2)
+ \nabla\cdot\Sp = -\mathbf{J}\cdot\mathbf{E}
= -\frac{\mathbf{v}}{c}\cdot (\mathbf{J}\times\mathbf{B}),
\end{equation}
where $\Sp$ is the Poynting flux, and we have used the fact that $B_0$
is constant to include it in the time derivative for future
convenience. Substituting into (\ref{eneqn1}) gives the equation for
the time-evolution of the non-thermal energy,
\begin{equation}
\label{eneqn2}
\frac{\partial}{\partial t}
\left(\frac{1}{2}\rho v^2 + \rho\phi + \frac{B^2-B_0^2}{8\pi}\right)
+\nabla\cdot \rho\mathbf{v} \left(\frac{1}{2}v^2+\phi\right)
+\nabla\cdot\Sp =
-\mathbf{v}\cdot\nabla P + \rho\frac{\partial\phi}{\partial t}
+\drho \left(\frac{1}{2}v^2 + \phi+\mathbf{v}\cdot\vkick\right).
\end{equation}

To include the internal energy, we write down the first law of
thermodynamics,
\begin{equation}
\rho \frac{de}{dt}+P\nabla\cdot\mathbf{v} = \Gamma - \Lambda,
\end{equation}
where $e$ is the internal energy per unit mass and $\Gamma$ and
$\Lambda$ are the rates of radiative energy gain and loss per unit
volume. Combining this with the continuity equation gives the
evolution equation for the internal energy,
\begin{equation}
\label{eneqn3}
\frac{\partial}{\partial t}\left(\rho e\right) +
\nabla\cdot\rho\mathbf{v} \left(e+\frac{P}{\rho}\right) =
\drho e + \mathbf{v}\cdot\nabla P + \Gamma - \Lambda.
\end{equation}

Adding together the evolution equations (\ref{eneqn2}) and
(\ref{eneqn3}) for the non-thermal and thermal energies gives the
total energy equation
\begin{eqnarray}
\lefteqn{
\frac{\partial}{\partial t}
\left[
\rho \left(\frac{1}{2}v^2+e+\phi\right)
+\frac{B^2-B_0^2}{8\pi}
\right]
+\nabla \cdot \rho\mathbf{v}
\left(\frac{1}{2}v^2 + e + \frac{P}{\rho} + \phi\right)
+ \nabla \cdot \Sp }
\nonumber \\
& \qquad \qquad \qquad \qquad = &
\rho \frac{\partial\phi}{\partial t}
+ \drho \left(\frac{1}{2}v^2+e+\phi+\mathbf{v}\cdot\vkick\right)
+ \Gamma - \Lambda.
\label{eneqn4}
\end{eqnarray}

To derive the global form of the energy equation, we integrate over
the virial volume $\Vv$, which gives
\begin{eqnarray}
\lefteqn{
\frac{d\cale}{dt} + \int_{\Sv} 
\left[\rho\left(\frac{1}{2} v^2 + e + \phi\right) + P\right]
\mathbf{v}\cdot d\mathbf{S} + \int_{\Sv} \Sp\cdot d\mathbf{S}
}
\nonumber \\
& = &
\frac{1}{2} \int_{\Vv} \rho \frac{\partial\phi}{\partial t} dV +
\frac{\dMc}{\Mc} (\cale - \calb)
+ \left(\frac{3-\krho}{4-\krho}\right)\dMc
\dRc v'_{\rm ej} + \calg_{\rm cl} - \call_{\rm cl},
\label{eneqn5}
\end{eqnarray}
where
\begin{equation}
\cale = \int_{\Vc} 
\left[\rho\left(\frac{1}{2} v^2 + e + \frac{1}{2}\phi\right) + 
\frac{B^2-B_0^2}{8\pi}
\right] dV
\end{equation}
is the total energy in the cloud, $\calg_{\rm cl}$ and $\call_{\rm
cl}$ are the total rates of radiative energy gain and loss integrated
over the cloud, and we have
used our assumption of homologous motion to evaluate the terms
involving $\drho$ and $\vkick$. To evaluate the first integral on the
left-hand side, we note all the terms proportional to density vanish
at the cloud surface because the ambient material has zero density,
but that constant pressure and zero density correspond to an
incompressible fluid, so the velocity of the ambient medium across the
virial surface is related to the velocity of the cloud surface by
$v=\dRc (\Rc^2/\Rv^2)$. This implies that
\begin{equation}
\label{presintegral}
\int_{\Sv} 
\left[\rho\left(\frac{1}{2} v^2 + e + \phi\right) + P\right]
\mathbf{v}\cdot d\mathbf{S} = 4\pi \Pa \Rc^2 \dRc.
\end{equation}
To evaluate the integral on the right-hand side, recall that gas in
the wind contributes negligibly to the gravitational potential. Since
the potential arises solely from cloud material, we can write
\citep{shu92}
\begin{equation}
\label{gravintegral}
\frac{1}{2} \int_{\Vv} \rho \frac{\partial\phi}{\partial t} dV
= \frac{1}{2} \int_{\Vv} \frac{\partial\rho}{\partial t} \phi \, dV
= \left(\frac{\dMc}{\Mc}\right) \frac{1}{2} \int_{\Vv} \rho \phi \, dV
= \frac{\dMc}{\Mc} \calw.
\end{equation}

Finally, we must evaluate $\int_{\Sv} \Sp\cdot d\mathbf{S}$, which
represents the flux of magnetic energy across the virial surface as it
is carried off by the wind. This is uncertain, because it depends on
the process by which the mass is removed. As in
\S~\ref{equationofmotion}, we
write the magnetic energy as a sum of turbulent and
non-turbulent contributions, $1/(8\pi) \int_{\Vv} (B^2-B_0^2)\,dV = 
\calb =
\calb_{\rm non-turb} + \calb_{\rm turb}$. For the turbulent part, we assume
that the wind does not change the nature of the MHD turbulence within
the cloud, so the turbulent magnetic energy remains proportional to
the turbulent kinetic energy. Since the loss of turbulent energy
accompanying mass loss is $\int_{\Vv} 3 \drho \sigma^2/2\,
dV = (\dMc/\Mc) \calt_{\rm turb}$, this requires that the flux of
turbulent magnetic energy be $\partial{\calb}_{\rm turb}/\partial t =
(\dMc/\Mc) \calb_{\rm turb}$. For the non-turbulent part, for simplicity and
based on the observation that GMCs have roughly constant mass-to-flux
ratios \citep{crutcher99}, we assume that the mass-to-flux ratio
remains constant as the cloud loses mass. Since $\calb_{\rm non-turb}
\propto \Phi^2$, this implies $\partial{\calb}_{\rm non-turb}/\partial t = 2
(\dot{\Phi}/\Phi) \calb_{\rm non-turb} = 2 (\dMc/\Mc) \calb_{\rm non-turb}$. Thus,
we can write
\begin{equation}
\label{magintegral}
\int_{\Sv} \Sp\cdot d\mathbf{S} = -\frac{\dMc}{\Mc} (\calb_{\rm turb} +
2 \calb_{\rm non-turb}) = -\frac{\dMc}{\Mc} (\calb - \etab^2 \calw).
\end{equation}

Substituting (\ref{presintegral}), (\ref{gravintegral}), and
(\ref{magintegral}) into (\ref{eneqn5}), we arrive at the final energy
equation for the cloud:
\begin{equation}
\label{energyeqn1}
\frac{d\cale}{dt} = \frac{\dMc}{\Mc} \left[\cale +
(1-\etab^2) \calw\right] - 4\pi\Pa\Rc^2\dRc +
\left(\frac{3-\krho}{4-\krho}\right) \dMc \dRc v'_{\rm ej} +
\calg_{\rm cl} - \call_{\rm cl}.
\end{equation}

\end{appendix}


\end{document}